\documentclass[a4paper,fleqn,usenatbib]{mnras}

\usepackage{mathptmx}

\usepackage[T1]{fontenc}
\usepackage{ae,aecompl}

\usepackage{graphicx}	
\usepackage{amsmath}	
\usepackage{amssymb}	
\usepackage{xspace}

\newcommand{\Lx}{$L_{\mathrm{X}}$\xspace}      
\newcommand{\chandra}{\textit{Chandra}\xspace}      
\newcommand{\rosat}{\textit{ROSAT}\xspace}      
\newcommand{\erg}{\,\mathrm{erg\, s^{-1}}}    
\newcommand{\gr}{\textit{g}--\textit{r}\xspace}    
\newcommand{\ri}{\textit{r}--\textit{i}\xspace}    
\newcommand{\iz}{\textit{i}--\textit{z}\xspace}    
\newcommand{\Mpc}{\ensuremath{\mathrm{Mpc}}}  

\DeclareGraphicsExtensions{.pdf, .png}     

\title[Recovering Clusters with AGN in Their Cores]{Hiding in Plain Sight - Recovering Clusters of Galaxies with the Strongest AGN in Their Cores}

\author[Green et al.]{
T. S. Green$^{1}$\thanks{E-mail:t.s.green@durham.ac.uk}, 
A. C. Edge$^{1}$,
H. Ebeling$^{2}$,
W. S. Burgett$^{3}$,
P. W. Draper$^{1}$,
N. Kaiser$^{2}$,
\and R.-P. Kudritzki$^{2}$,
 E. A. Magnier$^{2}$,
N. Metcalfe$^{1}$,
R. J. Wainscoat$^{2}$,
C. Waters$^{2}$
\\
$^{1}$Centre for Extragalactic Astronomy, Durham University, South Road, Durham DH1 3LE, UK\\
$^{2}$Institute for Astronomy, University of Hawaii at Manoa, Honolulu, HI 96822, USA\\
$^{3}$GMTO Corporation, 465 N. Halstead St., Suite 250, Pasadena, CA  91107, USA
}

\date{Accepted 2016 November 22. Received 2016 November 02 ; in original form 2016 September 12.}

\pubyear{2016}

\begin{document}
\label{firstpage}
\pagerange{\pageref{firstpage}--\pageref{lastpage}}
\maketitle

\begin{abstract}
A key challenge in understanding the feedback mechanism of AGN in Brightest Cluster Galaxies (BCGs) is the inherent rarity of catching an AGN during its strong outburst phase. This is exacerbated by the ambiguity of differentiating between AGN and clusters in X-ray observations. If there is evidence for an AGN then the X-ray emission is commonly assumed to be dominated by the AGN emission, introducing a selection effect against the detection of AGN in BCGs. In order to recover these `missing' clusters, we systematically investigate the colour--magnitude relation around some $\sim3500$ \textit{ROSAT} All Sky Survey selected AGN, looking for signs of a cluster red sequence. Amongst our 22 candidate systems, we independently rediscover several confirmed systems, where a strong AGN resides in a central galaxy. We compare the X-ray luminosity to red sequence richness distribution of our AGN candidate systems with that of a similarly selected comparison sample of $\sim1000$ confirmed clusters and identify seven `best' candidates (all of which are BL~Lac objects), where the X-ray flux is likely to be a comparable mix between cluster and AGN emission. We confirm that the colours of the red sequence are consistent with the redshift of the AGN, that the colours of the AGN host galaxy are consistent with AGN, and, by comparing their luminosities with those from our comparison clusters, confirm that the AGN hosts are consistent with BCGs.  
\end{abstract}




\section{Introduction}
\indent Feedback from Active Galactic Nuclei (AGN) activity in galaxies plays a crucial role in galaxy evolution, truncating star formation in massive galaxies \citep{Bower+06}. In the case of Brightest Cluster Galaxies (BCGs) their unique environment, at the centre of clusters of galaxies, leads to an interplay between AGN activity and the wider external environment of the galaxy, affecting the evolution of both the host galaxy and the surrounding intracluster medium (ICM).  

\indent The baryonic content of massive galaxy clusters is dominated by the ICM \citep{Lin+03}, a hot gas ($T\sim10^{7-8}\,\mathrm{K}$) which radiates in the X-rays, with typical luminosities of the order $\sim10^{43-45}\,\mathrm{erg\, s^{-1}}$. As energy is radiated away, a process of cooling occurs. So in cool core clusters -- that is, clusters with high central gas densities, hence highly peaked X-ray surface brightness profiles, and hence cooling timescales $\lesssim1$Gyr -- we naively expect to find large reserves of cool gas in their cores. In order to support the weight of the overlying gas, cooling flows of the order of $100\mathrm{s-}1000\mathrm{s}\,M_{\sun}\mathrm{yr}^{-1}$ were predicted \citep{Fabian94}, resulting in substantial rates of star formation in the BCG as the cool gas accumulates. However, high resolution X-ray observations have revealed a deficit in cool gas, with evidence for only $1-10\%$ of the predicted mass of gas cooling and going on to form stars (\citealt{O'dea+08}). 

\indent A heating mechanism therefore must exist to offset the radiative cooling, now widely accepted to be AGN feedback. The evidence points toward a self-regulated cycle, where gas cools, condenses into filaments (\citealt{Tremblay+12,McDonald+14}), leading to star formation (\citealt{Cavagnolo+08,Donahue+10,Rawle+12,Hoffer+12,Fogarty+15, Green+16}) and accretion onto the supermassive blackhole of the BCG. As the central black hole accretes matter, energy gets ejected back out into the system, heating the gas and halting the supply of gas for subsequent accretion and star formation. As the AGN `switches off', the process of radiative cooling takes over again. 

\indent This picture of a feedback cycle is primarily supported by observations of clear X-ray cavities surrounding BCGs with active nuclei (\citealt{Boehringer+93, McNamara+00, HL+15}), with cavities observed in $\geq70\%$ of cool core clusters (\citealt{Dunn+06,Hlavacek+12}). Cavities are spatially coincident with positions of radio lobes created by the AGN (with the majority of cool core clusters observed to have a central radio galaxy; \citealt{Burns+90,Hogan+15a}) and are believed to be driven by the outflowing plasma. The AGN power, estimated by measuring the mechanical power necessary to create the cavities, is found to be generally sufficient to offset the rate of radiative cooling in most cool core clusters (\citealt{Birzan+04, Rafferty+06}). Although it should be noted that the high variability of AGN power (\citealt{Hogan+15b}) introduces a further challenge when comparing AGN power with cooling rates.
However, the precise nature of how the energy gets transferred is still a topic of uncertainty and debate (see reviews \citealt{McNamara+07, McNamara+12, Fabian12}).

\indent A key difficulty in understanding this feedback cycle is the extreme rarity of catching BCGs during their active outburst stage. Due to the short duty cycle of AGN, such systems are inherently rare, but this is further exacerbated by observational selection effects. The principle challenge is the ambiguity that exists between AGN and cluster emission in X-ray observations, i.e. whether X-rays are associated with extended cluster emission or point-like emission from AGN. For example, cool core clusters, the most likely systems to host a BCG with an active nuclei, can appear as point sources in shallow X-ray data, due to their centrally peaked profile, and as a result be ignored \citep{Pesce+90}. The effect particularly applies for cool core clusters at high redshift and may have a contribution towards the apparent decrease in the cool cores fraction with redshift (\citealt{Vikhlinin+07,Santos+08}). This is important to bear in mind in future  X-ray surveys, such as  \textit{eROSITA}, as the extent of a cool core X-ray peak profile can be as little as 10 arcsec at high redshift ($z\gtrsim0.5$), and hence would be unresolvable with the effective PSF of 28 arcsec in \textit{eROSITA}'s scanning mode. Additionally, moderately X-ray luminous AGN in clusters are difficult to detect above the cluster X-ray emission of the cluster core and AGN in clusters often lack optical spectral tracers of AGN activity \citep{Martini+06}. Conversely, strong AGN can have all the X-ray flux attributed to them, whilst the cluster they may reside in goes uncatalogued. In this paper we attempt to address the latter issue.
 
\indent When compiling an X-ray catalogue of clusters, it is common practice to classify an X-ray source as an AGN whenever evidence, such as strong high-ionisation emission, suggests the presence of nuclear activity. This classification would remain unchanged even in the obvious presence of a galaxy overdensity in the direction of the source, since probabilistic arguments strongly favour QSO dominance of the X-ray emission in the case of a cluster / QSO chance alignment \citep{Zenn+10}.
However, if the AGN host galaxy is a member of a massive galaxy cluster the X-ray emission from the AGN may be a minority contributor to the total combined AGN and cluster emission. As a result, there is an observational selection effect against detection of rich clusters with an AGN in the BCG - the very sources we need in order to better understand the feedback cycle in these systems.

\indent A case in point is the Phoenix cluster, SPT-J2344-42 \citep{McDonald+12}. Despite being one of the most X-ray luminous clusters known and one of the $\sim2000$ brightest \rosat All Sky Survey sources (RBS2043, \citealt{Schwope+00}), this cluster went undetected for a long time. The X-rays were attributed to AGN emission due to the strong, broad Mg II line in the spectrum of the BCG. Only when it was later detected through the S-Z effect, via the SPT, was it realised to be a massive cluster. A follow-up \chandra observation revealed the cluster emission ($6\times10^{45}\erg$) to be 3 times that of the AGN ($2\times10^{45}\erg$; \citealt{McDonald+12}), and a subsequent \textit{HST} observation revealed blue filaments in the BCG and a star formation rate of $748M_{\sun}\,\mathrm{yr}^{-1}$ \citep{McDonald+13}.    

\indent One additional source of confusion in the literature is that some 
cluster-dominated X-ray sources are identified as candidate BL~Lac objects. 
BL~Lac objects (reviews: \citealt{Stein+76,Falomo+14}) are a class of AGN blazars, 
with emission over the entire electromagnetic spectra, 
due to non-thermal processes in a 
relativistic jet, orientated to be near the line of sight (\citealt{Urry+95}). They are characterised by featureless optical spectra, so detection is usually made via X-ray, radio and $\gamma$-ray detections. 
As a result some BCGs which exhibit a relatively
bright, flat-spectrum radio source are identified as BL~Lacs when the cluster is in fact the dominate 
source of X-rays. For example, at least 40 systems are included in the Roma BZCAT sample (the largest sample of BL~Lac objects compiled to date, containing over 3500 blazar objects; \citealt{Massaro+09}), where the X-ray 
emission is likely to be dominated by the host cluster (see Section~\ref{sect:roma} for details). 
Similarly, seven of the 
targets chosen for OVRO~40m monitoring as potential {\it Fermi} sources are in
fact BCGs (\citealt{Hogan+15b}). While this has a relatively minor effect on these
large samples, it is important to ensure that the correct associations are made
in order to avoid the clusters with a BCG, in a rare episode of AGN activity, being
lost from an X-ray flux-limited sample. We will return to this issue in
the discussion.

\indent The reverse is also possible however, in which low luminosity BL~Lacs can
be misidentified as clusters of galaxies \citep{Rector+99}. While
relatively few clusters have been identified in subsequent X-ray
pointed observations to be AGN-dominated, there are several
notable cases of systems where the 
cluster is a minority contributor to the total X-ray
emission, for example: A1366 and A2627 in \cite{Russell+13}, A1774 and A2055 (\citealt{Ebeling+96})
 and the well studied samples
of 1H1821+64 and 3C186 (\citealt{Hall+97, Siemiginowska+10}). 
In these cases, it is very important to 
pay attention to the emission of both the cluster \textit{and} the AGN,
in order to understand the total mass of the cluster and hence
the parent sample of possible clusters that could have
hosted a particular AGN.
 
\indent The primary goal of this paper is to perform a systematic search for signs of a rich cluster around \rosat X-ray sources identified as AGN, with the aim of recovering these `missing' clusters with strong AGN in their cores, i.e. Phoenix cluster analogues. In order to do this we investigate the colour-magnitude relationship around the AGN and attempt to look for signs of a overdensity in red sequence galaxies. Specifically, we look for cluster red sequences with a richness comparable to those of confirmed clusters of similar X-ray luminosity to our ``AGN'' X-ray sources.  

\indent The organisation of this paper is as follows: we introduce the AGN sample and the cluster comparison sample in Section \ref{sect:data}. We describe our photometry and red sequence selection methodology in in Section \ref{sect:method}. We present the results and discuss these in Section \ref{sect:results} and conclude with a summary in Section \ref{sect:conclude}. Throughout this paper optical photometry is given in AB magnitudes, and \textit{WISE} photometry in Vega. A standard cosmology of $H_0 = 72\, \mathrm{km\,s^{-1}\,Mpc^{-1}}$, $\Omega_M = 0.27$ and $\Omega_{\Lambda}=0.73$ is assumed.

\section{The Data}\label{sect:data}

\subsection{AGN Sample}
Our parent AGN sample consists of all X-ray sources from the \rosat All-Sky Survey (RASS) Bright Source Catalogue \citep{Voges+99}, which were attributed to AGN emission through the combined follow-up over the past 20 years - including large spectroscopic surveys such as SDSS (\citealt{Anderson+03, Anderson+07}), and 6dFGS (\citealt{Mahony+10}), plus targeted projects (\citealt{Bauer+00, Kollatschny+08}). This provides a comprehensive catalogue of AGN to a soft X-ray flux limit of 3--5$\times 10^{-13}$~erg~s$^{-1}$~cm$^{-2}$ at 0.1--2.4~keV. Within the PS1 3$\pi$ footprint, and below a redshift of 0.4, this gives a sample of 3058 broad-line AGN and 412 BL~Lacs. The X-ray luminosities are all drawn from the RASS, are quoted in the 0.1--2.4~keV band and corrected for Galactic absorption. 

There are a further 1020 RASS BSC sources, likely to be AGN, within the PS1 3$\pi$ footprint, which have not been spectroscopically confirmed. However, we restrict our analysis to spectroscopically confirmed AGN only, as without the redshift -- red sequence colour relationship (see Section~\ref{Sect:rscolz}) we would not be able to verify redshift concordance between the AGN and any apparent over-density. As our working AGN sample consists of a total of 3470 spectroscopically confirmed BL~Lac and broad-line AGN, our spectroscopic completeness is at least 77\%. In all likelihood however the completeness is higher than this for the brighter optical counterparts, as the fainter AGN without spectroscopy are more likely to be at a redshift above our $z=0.4$ cut-off.

\subsection{Comparison Cluster Sample}
In order to check for consistency between candidate systems and clusters we compare our observations with an X-ray selected sample of 981 confirmed clusters. The cluster sample, which covers $0.03<z<0.5$, was drawn from a systematic investigation into the RASS Bright Source Catalogue \citep{Voges+99}, so has a similar selection criteria to our AGN. This is inclusive of all published samples within the PS1 $3\pi$ footprint, such as the Brightest Cluster Sample, BCS \citep{Ebeling+98}, the extended Brightest Cluster Sample, eBCS \citep{Ebeling+00}, the ROSAT ESO Flux Limited X-ray survey, REFLEX \citep{Bohringer+04}, the Northern ROSAT All-Sky Galaxy Cluster Survey, NORAS, \citep{Bohringer+00} and the Massive Cluster Survey, MACS, (\citealt{MACS, Ebeling+07, Ebeling+10}). In addition to these, the sample also includes all cluster identifications of RASS Bright Source Catalogue sources below the flux limits of these published surveys. The X-ray luminosities are all drawn from the RASS and are quoted in the 0.1--2.4~keV band and corrected for Galactic absorption. In \cite{Green+16} this cluster sample was used to investigate activity in BCGs through multi-wavelength photometry; we refer the reader to this paper for details regarding the computation and distribution of X-ray luminosities for these clusters. .

\section{Methods}\label{sect:method}

\subsection{Optical Photometry}
For the optical analysis we utilised the Pan-STARRS, PS1 $3\pi$ survey \citep{Tonry+12}, which is an optical wide-field photometric survey covering the entire sky north of a declination of $-30\degr$. The survey covers the $g,r,i,z$ and $y$ bands at an imaging spatial resolution of 0.25 arcsec per pixel. For each \rosat AGN we extracted the PS1 3$\pi$ PV3 postage stamp and performed aperture photometry in the \textit{griz} bands using SExtractor \citep{Bertin+96}. The exact same procedure for the PS1 3$\pi$ photometry, and settings in SExtractor, were used for the AGN sample as for the cluster comparison sample, (as outlined in more detail in \citealt{Green+16}). Specifically, we determine magnitudes using the MAG\_AUTO parameter, which measures the flux within a flexible elliptical aperture with a Kron radius \citep{Kron80}, and colours are derived from the MAG\_APER parameter, with a fixed circular aperture of 4 arcsec. The inbuilt CLASS\_STAR function was used for star-galaxy separation. All PS1 photometry throughout this paper is corrected for Galactic reddening using the Galactic Extinction calculator available through the NED, based on the \cite{Schlafly+11} extinction maps. A K-correction is applied, when determining optical luminosities, assuming a simple stellar population model, \citep{Bruzual+03}, with solar metallicity, a Chabrier Initial Mass Function, formation at $z=3$ and subsequent passive evolution.

\subsection{Mid-IR Photometry}
For the Mid-IR analysis we utilised the \textit{Wide-field Infrared Survey Explorer} (\textit{WISE}) \citep{Wright+10}, which is a survey covering the whole sky at $3.4$--$22\,\micron$. From the AllWISE Source Catalogue we extract photometry for our AGN/BCG sample in the \textit{W}1, \textit{W}2 and \textit{W}3 bands at $3.4$, $4.6$ and $12\,\micron$ respectively. We do not perform any Galactic extinction corrections as this is negligible for Mid-IR observations \citep{Cardelli+89}. 

\subsection{Red Sequence Selection}

\indent We extract the photometry of sources within a $0.5\,\Mpc$ radius about the X-ray position, and create \gr, \ri and \iz colour-magnitude diagrams for each AGN position. The colour-magnitude diagrams were then visually inspected for signs of a colour--magnitude relation consistent with the red sequence seen in all rich clusters of galaxies. A red sequence selection algorithm was applied to any X-ray sources with a visually apparent red sequence. The algorithm is described in more detail in \cite{Green+16}, but briefly it applies an initial best fitting line to an apparent red sequence. Then, in decreasing intervals of colour around the line, a new best fitting line is made until finally sources within $\pm0.1$ mag around the line are selected. All the while, the fitting is only made to sources which lie on the red sequence in two separate colour indices, where the most blue-ward index spans the $4000$\r{A} break (e.g. \gr and \ri at $z<0.375$, and \ri and \iz at $z>0.375$).

\indent For X-ray sources where an apparent red sequence exists, the PS1 imaging was then visually inspected in order to (a) check for a galactic over-density that is visually consistent with that of a cluster, and (b) check if the AGN host was visually consistent with being a BCG -- that is, at the centre of the galactic distribution of the cluster and preferably with an extended cD-like envelope. Any sources satisfying these criteria have been identified as candidate systems in the subsequent analysis. Note that in addition to these there were a number of the AGN which showed a red sequence in their colour-magnitude diagram, and hence are likely to be hosted by a cluster member, but were clearly not the central galaxy, and hence not the focus of this analysis.

\section{Results} \label{sect:results} 

Of the 3058 broad-line AGN and 412 BL~Lac the selection was narrowed down to 22 candidates. These are AGN which satisfy the conditions of (a) exhibiting a colour--magnitude diagram consistent with a cluster red sequence and (b) are hosted by a galaxy visually identifiable as a BCG. Interestingly we find these are almost exclusively BL~Lac objects (20/22), with only two of the broad-line AGN satisfying our selection criteria. This suggests broad-line AGN are a very rare episode in the lifetime of any given system.
 
\indent Our confidence in the technique used is strongly supported by the independent rediscovery of several previously identified systems where a BCG hosts a strong AGN. Some of which are well known; 1H0414+11, 1H1821+644, A2055, RXJ0910+3106, RXJ1745+3951. These five systems have had subsequent pointed X-ray observations confirming the presence of extended cluster emission in addition to the AGN. These observations allow the relative X-ray contributions to be estimated, providing context for our new discoveries - details of which are given in Section~\ref{sect:rediscover}. The independent rediscovery of these clusters, which represent the very systems we are aiming to uncover, strongly supports the validity of our approach.

\subsection{Cluster Membership}\label{Sect:rscolz}
In order to reduce chance projection effects along the line of sight, between a cluster and an unassociated AGN, we investigate whether the AGN is consistent with cluster membership. In lieu of having the required spectroscopy to spectroscopically confirm cluster members, we utilise the relationship between observed red sequence colour and redshift for clusters (\citealt{Andreon03, Rykoff+14, Green+16}). In \cite{Green+16} it was shown that the cluster redshift of our comparison cluster sample can be photometrically estimated to within $\Delta_z=0.025$, at $1\sigma$, using the \gr red sequence colour. 

\begin{figure}
	\includegraphics[width=1\columnwidth]{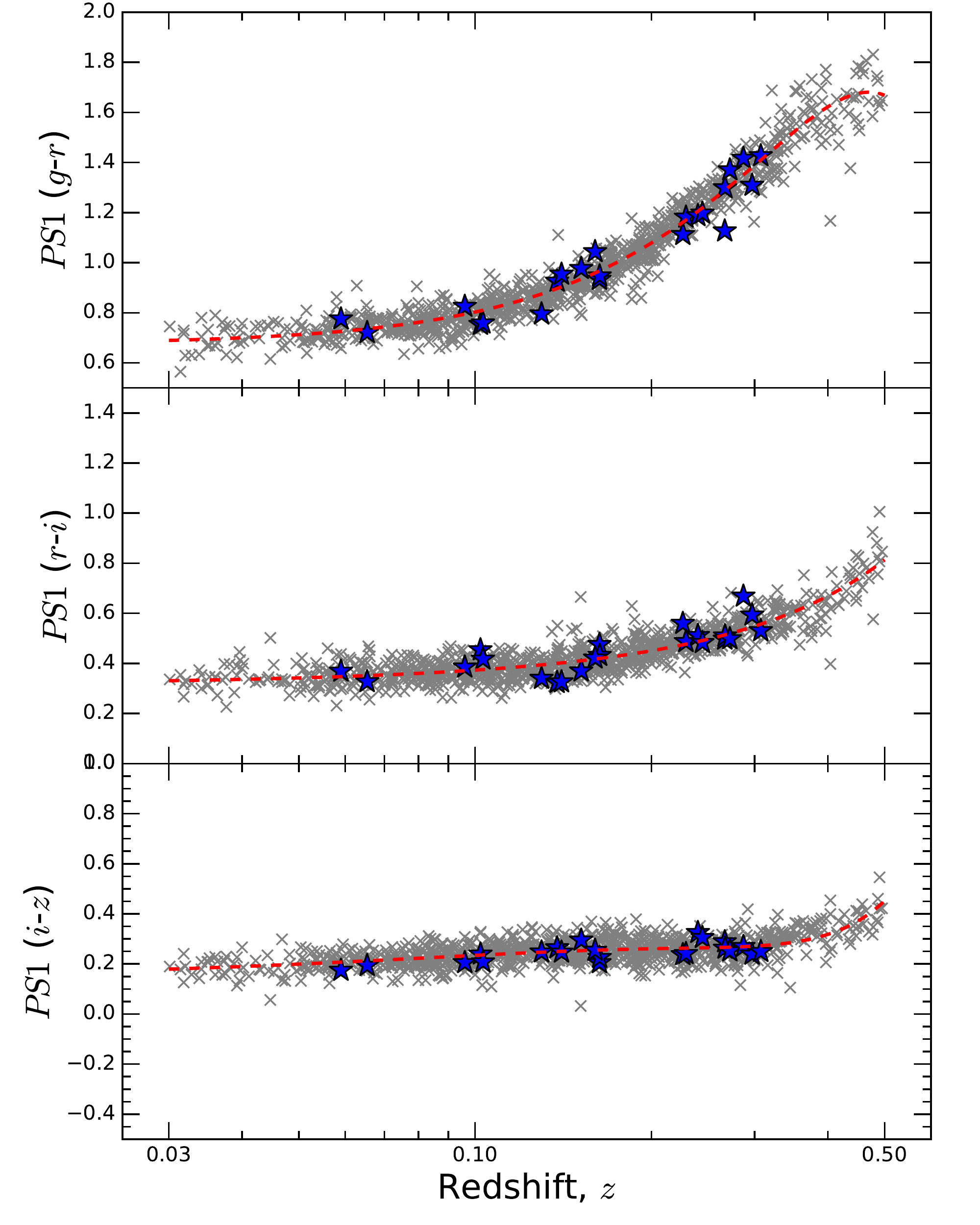}
    \caption{The PS1 \gr, \ri and \iz colour of the red sequence, measured at 19th magnitude, against redshift. The (grey) crosses represent the red sequence of each of the confirmed clusters from our cluster comparison sample. The (blue) stars represent the red sequence around the AGN from this work. The (red) dashed line is a best fitting line to the comparison cluster data. All of our candidate systems have a red sequence colour consistent with the given redshift of the AGN.}
    \label{fig:RScol-z}
\end{figure}

\indent In Fig. \ref{fig:RScol-z} we overlay the colour of the apparent red sequences, selected around our AGN, on top of that from our cluster comparison sample. Red sequence colour is defined as the colour, given by the best fitting line of the red sequence, at a flux of 19th magnitude. We find in all of our candidate systems that the red sequence colour is consistent with the colour predicted by the confirmed redshift of the AGN. This suggests that the galactic overdensity we detect is likely at the same redshift of the AGN and hence that the host galaxy is likely a cluster member. 

\begin{figure*}
	\includegraphics[width=1.5\columnwidth]{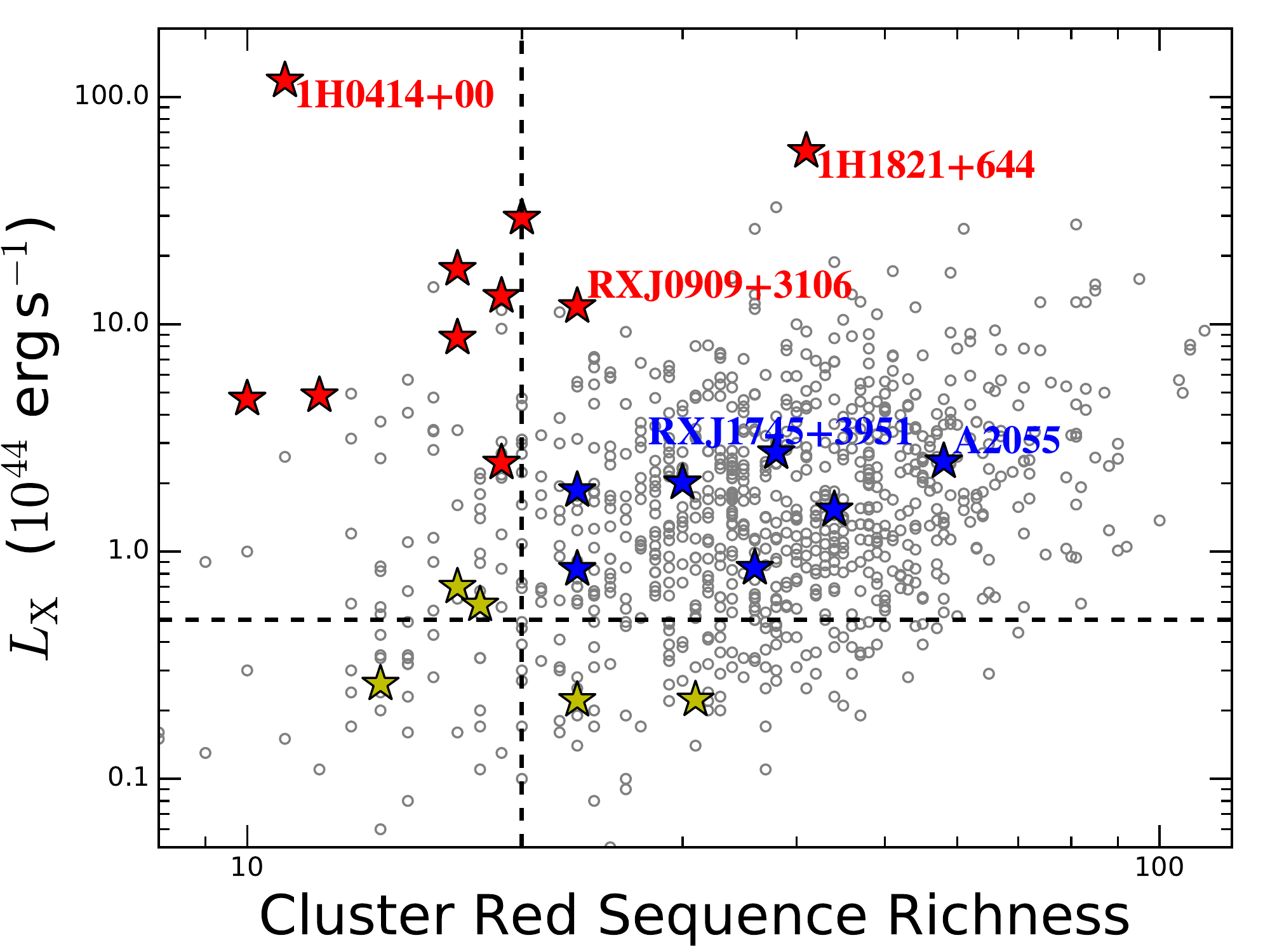}
        \caption{\rosat All Sky-Survey X-ray luminosity against red sequence richness (number of red sequence galaxies above $i$-band luminosity of -19 mag). The (grey) empty circles represent the comparison cluster sample. The coloured stars represent the 22 candidate systems where we suspect the presence of a cluster around the AGN. The stars with IDs represent the previously confirmed systems which we have independently rediscovered. The different coloured stars represent distinct groupings on the plot, which are used as reference in the text, (i.e. red = high \Lx for relative richness, suggesting probable X-ray point source dominance, yellow = relatively low \Lx and richness, suggesting AGN in group scale environments). The blue stars represent our `best' candidates as they best match the \Lx--richness distribution of our confirmed comparison sample of clusters. The `best' systems are those most likely to constitute a strong AGN in the BCG of a rich cluster, where the X-rays are a comparable mix of cluster and AGN emission. The dashed lines indicate the selection criteria for these `best' candidates.  }
    \label{fig:richness-Lx}
\end{figure*}
\subsection{Red Sequence Richness}
The optical richness of a galaxy cluster, i.e. the number of galaxies in it, scales weakly with the cluster mass (\citealt{Hall+97, Johnston+07, Andreon+10}). Hence we expect that the richness of the cluster red sequence -- that is, the number of galaxies selected on the cluster red sequence -- should scale with the X-ray luminosity of the cluster, (since \Lx traces cluster mass). In Figure \ref{fig:richness-Lx} we plot the X-ray luminosity against the red sequence richness of our cluster comparison cluster sample, where richness is defined as the number of red sequence galaxies brighter than an \textit{i}-band luminosity of -19 magnitudes, (corresponding to a flux of 16.5 mag at $z=0.03$ and 22.6 mag at $z=0.4$). We find a weak correlation between \Lx and red sequence richness for the cluster sample, but with considerable scatter. A source of scatter in the richness is that we perform our optical photometry within a fixed $0.5\,\Mpc$ aperture around the central galaxy. Hence, richness is actually a measure of galaxy density, so diffuse clusters, which tend to be unrelaxed systems that are currently undergoing, or have recently undergone, mergers will be measured with lower richness. So while for any given system one can not assume that because a cluster has a high X-ray luminosity then it will be have a substantially rich red sequence, the overall distribution does provide valuable context for our candidate AGN in cluster cores. 

\indent Applying the same procedure of measuring the richness to the 22 AGN candidate systems, we can investigate how the richness of our AGN colour--magnitude relations compare to those of confirmed clusters of similar X-ray luminosity (Fig. \ref{fig:richness-Lx}). By selection, we find all the candidate AGN lie within the scatter of the confirmed cluster sample in terms of richness. So, we can be reasonably confident that all these candidates represent a system where an AGN resides in the central galaxy of at least a group scale structure. 

\indent The candidate systems can roughly be divided into three distinct regions, which we colour code for clarity. The systems in the top left (red stars in Fig. \ref{fig:richness-Lx}), which have high X-ray luminosities relative to their richness, are most likely systems where the AGN dominates the X-ray emission. We identify two well known sources within this selection, 1H0414+00 and 1H1821+644, where the AGN is known to dominate. With the exception of 1H1821+644, these tend to have a low richness, suggesting they are likely low mass cluster/group scale overdensities. However it is possible that some are analogues to A689, a system in which a BL~Lac is known to reside in a cluster core (although not in the BCG itself) and significantly contaminates the X-ray emission. A \textit{Chandra} observation revealed an X-ray luminosity for the cluster as $2.8\times10^{44}\erg$ when the BL~Lac is masked \citep{Giles+12}, an order of magnitude less than the original BCS value of $3.0\times10^{45}\erg$. However this still represents a substantial X-ray luminosity, with a sufficient flux to be detectable as an X-ray cluster in the RASS data. Applying our red sequence selection algorithm yields a richness of 25 for this cluster, making its position in richness--\Lx comparable to RXJ0909.9+3106, i.e. over luminous for its given richness. This supports the interpretation that these represent systems where the AGN contribution dominates by as much as an order of magnitude over that of the cluster. Details on these systems are given in Table~\ref{Table:redstars}. The candidate systems in the lower left (yellow stars), with a low \Lx and/or richness, are likely systems with a moderate AGN in a low mass cluster/group scale system. Details on these systems are given in Table~\ref{Table:othercandidates}.

\indent It is the candidates in the centre however, (blue stars), which best match the richness--\Lx distribution of our confirmed clusters. As such, they are the systems most likely to exhibit a comparable mix of X-ray emission from the AGN and a rich cluster component, analogous to the Phoenix cluster. We designate these systems, with a richness in excess of 22 red sequence galaxies and \Lx$>0.5\times 10^{44}\erg$, (the dashed lines in Fig. \ref{fig:richness-Lx}), as our `best' candidates. This consists of seven systems, two of which are previously identified systems (A2055 and RXJ1745+3951; see section \ref{sect:rediscover}). Details on these systems are given in Table~\ref{Table:bestcandidates}. 

We note there is a more ambiguous source, RXJ0828.2+4154, with a richness narrowly below our best candidate selection criteria, but with a comparable \Lx to confirmed clusters (richness$=19$ and \Lx$=2.46\times10^{44}\erg$). However, from a \textit{Chandra} observation, this source appears to be point source dominated (Donahue private communication). 

\indent Note, in the appendix (Fig.\ref{fig:CMR}), the colour--magnitude diagrams and $gri$ colour composite images of the cluster and of the BCG, for our `best' candidates and previously confirmed systems, are provided.
 
\subsection{BCG/AGN Luminosity}
\begin{figure}
	\includegraphics[width=1\columnwidth]{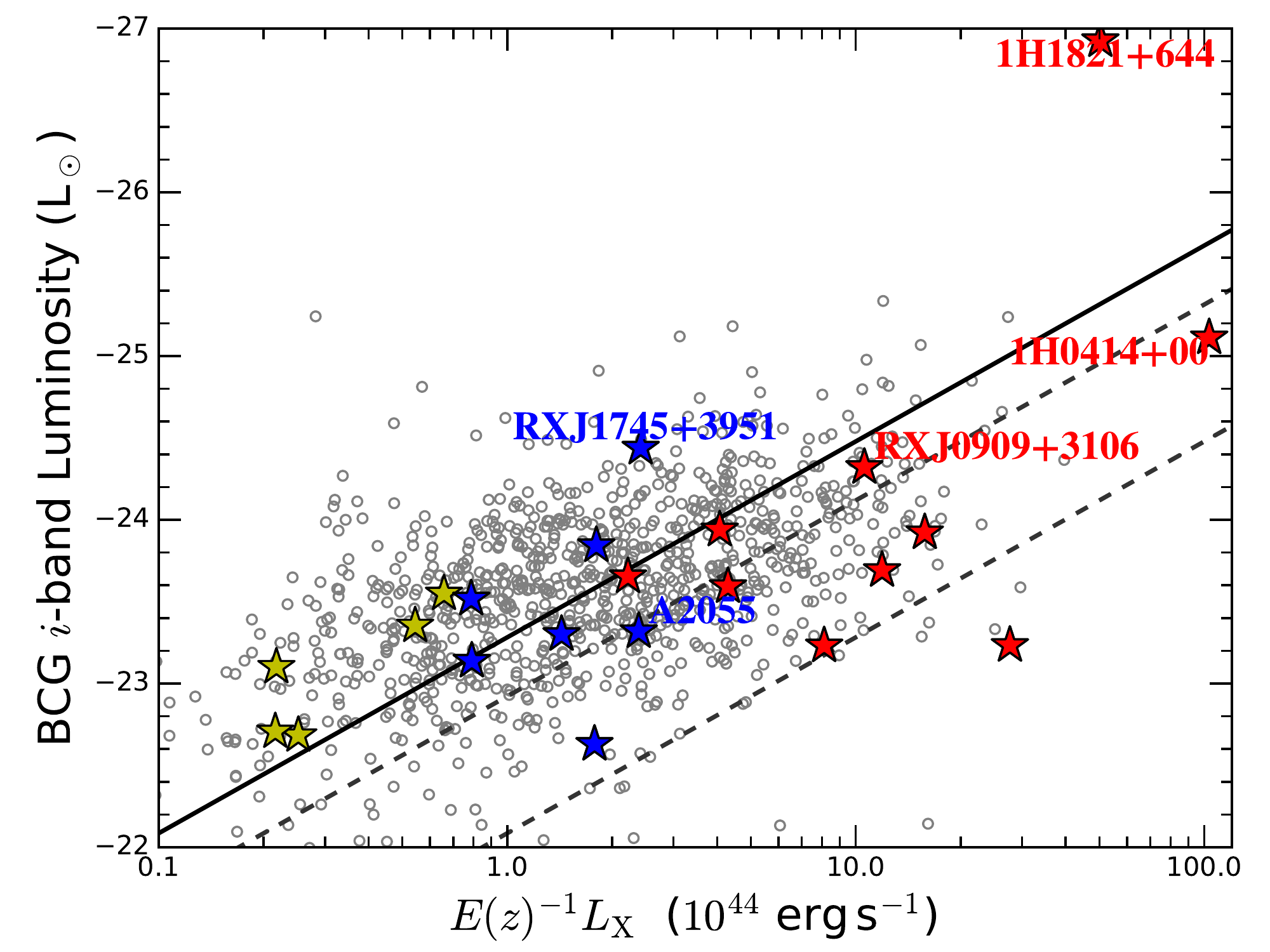}
    \caption{PS1 \textit{i}-band luminosity against X-ray luminosity. The open (grey) circles represent the BCGs of the comparison cluster sample. The coloured stars, (where colours follow from Fig.~\ref{fig:richness-Lx}), represent the host galaxy of the AGN. The solid line indicates the best fitting line to the cluster sample. The dashed lines show the best fitting line offset by a factor of two or ten respectively on the \Lx axis. So, for example, the outer dashed line indicates the best fitting relationship of $L_{i}$--\Lx for a system where \Lx is increased by a factor ten, (such as if a strong AGN were contributing X-rays at a ratio of 9:1 to the X-rays from the cluster).}
    \label{fig:BCGi-Lx}
\end{figure}

\begin{figure}
	\includegraphics[width=1\columnwidth]{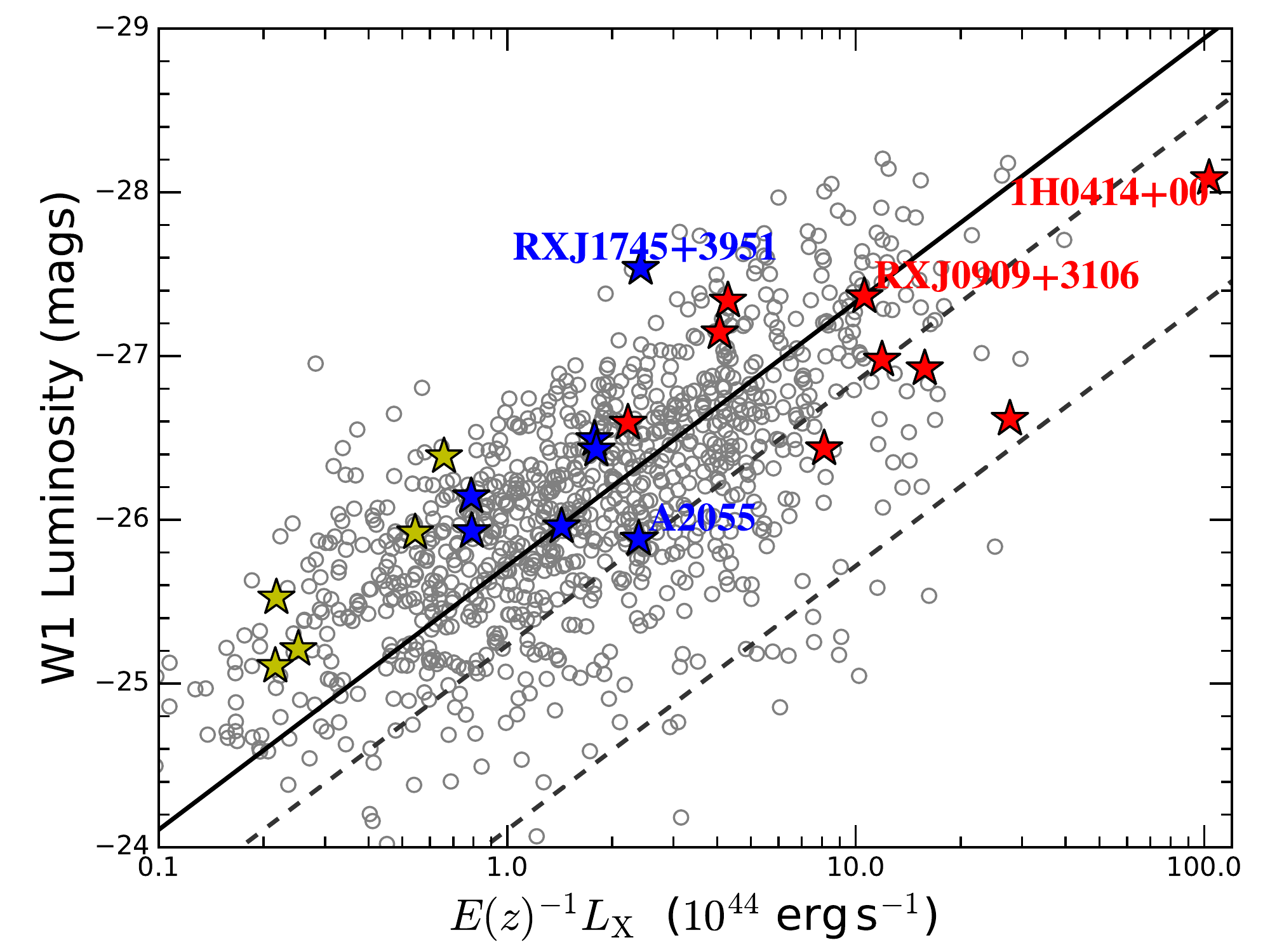}
        \caption{Similar to Fig. \ref{fig:BCGi-Lx} but with \textit{WISE} W1 ($3.6\micron$) luminosity against X-ray luminosity. Note, there is an additional data point at $L_{W1} = −31.14$ and \Lx$=57.8$, corresponding to 1H1821+644, which is omitted to maintain visual clarity.
}
    \label{fig:Lw1-Lx}
\end{figure}

The luminosity of BCGs has been shown to correlate with the X-ray luminosity of their host cluster (e.g. \citealt{Edge91, Stott+12, Green+16}). Hence in order to check whether our AGN host galaxies are consistent with being BCGs we check the luminosity of the AGN host galaxies in both the optical and Mid-IR. In the PS1 $i$-band we find the AGN host galaxies have luminosities generally consistent with those of BCGs of similar X-ray luminosity clusters (Fig. \ref{fig:BCGi-Lx}) - indicating these host galaxies have stellar masses consistent with those of BCGs. Similarly in the $W1$-band, at $3.6\micron$, the luminosities of the AGN hosts are generally consistent with BCGs of similar \Lx (Fig. \ref{fig:Lw1-Lx}).

A challenge in the interpretation of these plots is that the AGN itself can contribute to the luminosity of the host BCG, as well as to the X-rays. Hence we maintain the colour coding utilised in Fig. \ref{fig:richness-Lx} as their relative positions in richness--\Lx space provide context in addressing this degeneracy. For example, we notice the majority of the red starred candidate systems appear under-luminous in $i$, with respect to \Lx, suggesting either they are dimmer in $i$ than most BCGs or they have a higher \Lx than similar clusters. But, given these sources have a low richness for their given \Lx, it is likely the enhancement of the X-rays, due to the dominant nature of the AGN, is the most significant factor here. 1H1821+644 is over an order of magnitude brighter, in both $i$ and \textit{W}1, than the most luminous BCG from our comparison sample, indicating the luminous quasar dominates the optical and MIR luminosities.

\indent Our best candidate systems are generally well centred with respect to the confirmed BCG sample. However one of our `best' candidate systems also appears under-luminous in $L_i$, RXJ0056.3-0936. However, this particular system appears to be a multi-component, dumbbell BCG (Fig. \ref{fig:CMR}). In this dumbbell system, the northern component is more consistent with being the BCG, (with $L_{i} = -23.3$, placing it securely within the main distribution of BCGs in Fig. \ref{fig:BCGi-Lx}), and the BL~Lac object is in the southern component. While projection effects within the cluster can not be ruled out, we identify this system as a candidate because it is likely the two components are in the early stages of a merger. This assumption is based on the findings of a follow up study with MUSE IFU spectroscopy of dumbbell BCGs, from the cluster comparison sample, which reveal the majority of dumbbell systems have kinematics consistent with the two components being bound (Green et al. in prep.). Note, as the \textit{WISE} resolution is insufficient to resolve the two separate components of this dumbbell, the $L_{W1}$ of this galaxy is consistent with the comparison BCGs. 

\subsection{Confirmation of AGN Nature: BCG/AGN Colours}
One potential concern is that some of the sources may actually just be `ordinary' clusters -- that is a cluster without an AGN in the BCG -- which were simply misidentified in the initial RASS X-ray analysis. With this in mind, we investigate the colours of the AGN host galaxies in order to look for properties consistent with AGN. In \cite{Green+16} we measured the colours of the BCGs of the cluster comparison sample and found the bulk population of BCGs were passively evolving. However we found that $\gtrsim14\%$ showed significant colour offsets from passivity due to star formation and/or AGN activity in the BCG. We expect therefore that the AGN in this study should show significant offsets in colour with respect to that of a passive BCG. In addition to investigating the colours, we also check the NVSS data and find that all of our sources are detected at 1.4 GHz, another indicator of the active nature of these galaxies. 

\begin{figure}
	\includegraphics[width=1\columnwidth]{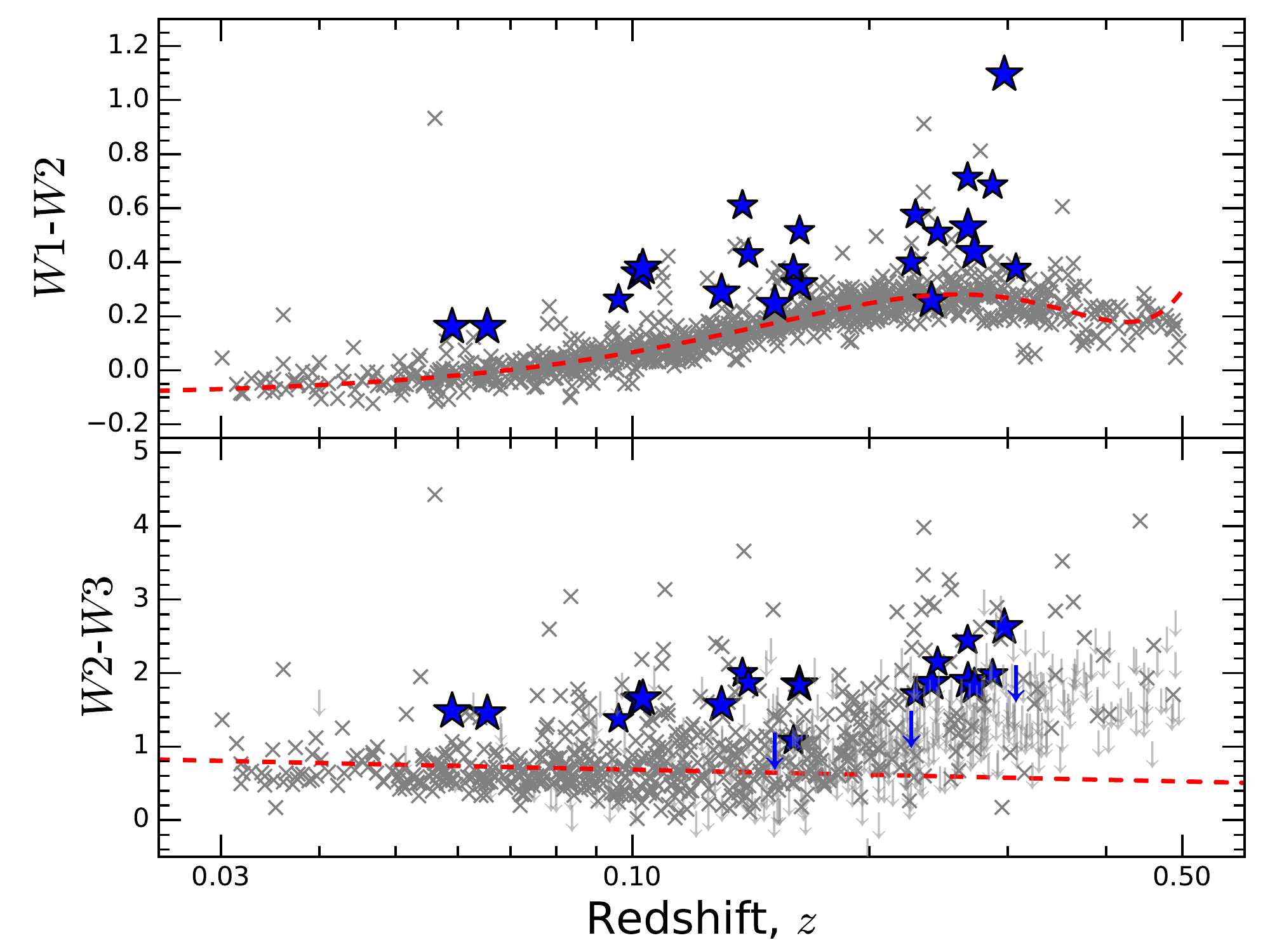}
    \caption{\textit{WISE} colours against redshift. The grey crosses represent the BCGs of our cluster comparison sample. The (blue) stars represent the AGN host galaxies. Sources with a \textit{W}3 signal to noise ratio, S/N$<3$, are indicated by arrows.}
    \label{fig:wise-z}
\end{figure}

\indent In the MIR star formation/AGN activity is characterised by enhanced emission at longer wavelengths, due to reprocessed dust emission, hence our AGN should be redder than passively evolving galaxies of similar redshift. In \textit{WISE} observations we expect AGN to be red in both \textit{W}1--\textit{W}2 and \textit{W}2--\textit{W}3 but star forming galaxies to be red in just \textit{W}2--\textit{W}3. In Fig. \ref{fig:wise-z} we overlay the \textit{WISE} \textit{W}1--\textit{W}2 and \textit{W}2--\textit{W}3 colours for our AGN host galaxies over those of the comparison BCG sample and find that almost all of our AGN candidates are red relative to passive BCGs in both colour indices. There is one source, RXJ2353.4-1257, which appears not to show the expected $4.5\micron$ excess, but a archival NTT spectra is consistent with the BL~Lac identification. This supports the assumption that these candidate systems are AGN, rather than a misidentified `ordinary' clusters.   

\begin{figure}
	\includegraphics[width=1\columnwidth]{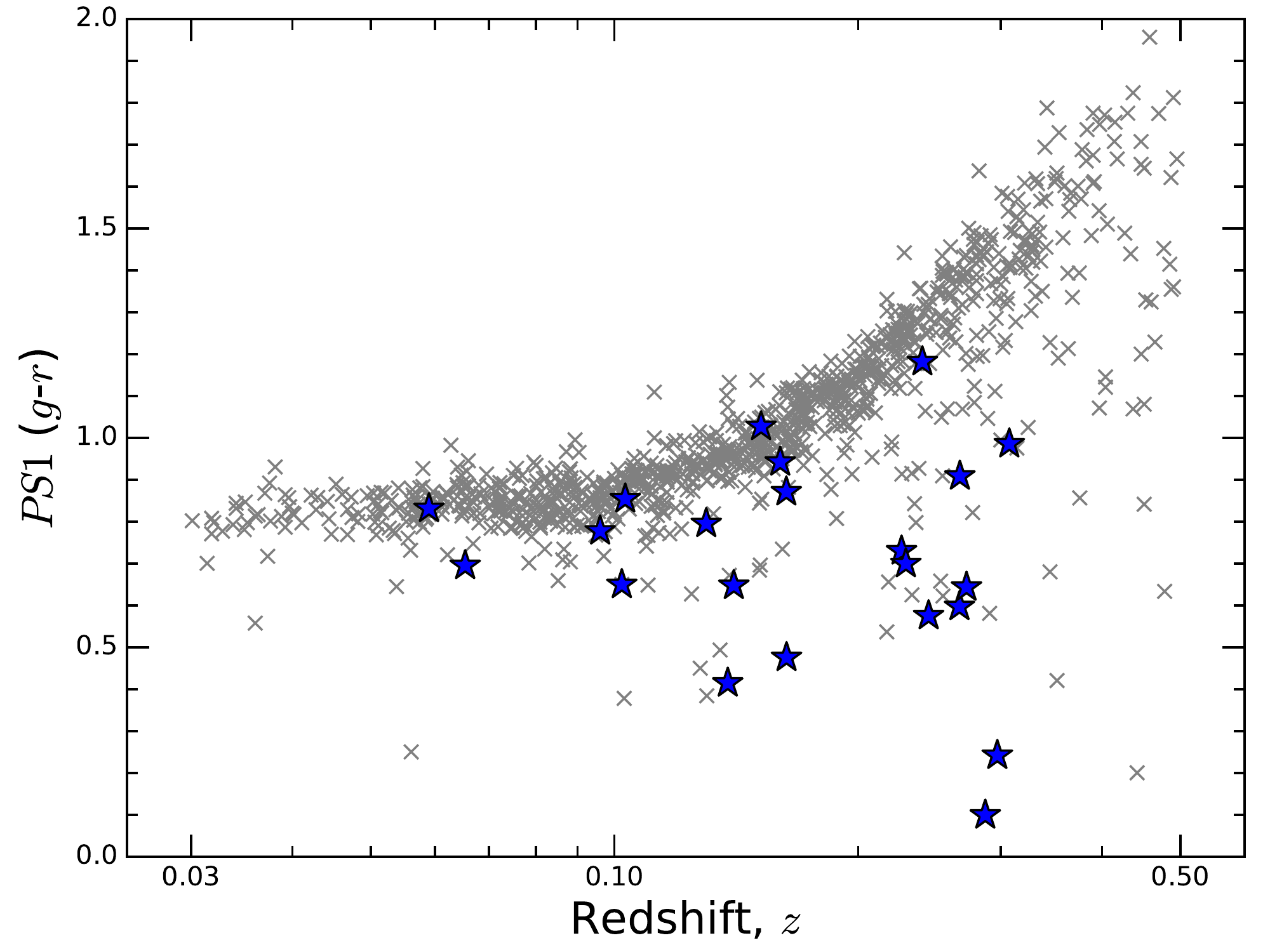}
    \caption{PS1 \textit{g}--\textit{r} colours against redshift. The grey crosses represent the BCGs of our cluster comparison sample. The (blue) stars represent the AGN host galaxies.}
    \label{fig:gr-z}
\end{figure}

\indent In Fig. \ref{fig:gr-z} we overlay the PS1 $3\pi$ \textit{g}--\textit{r} colour of our AGN host galaxies over those of the comparison BCGs. We find that the majority of our AGN hosts are significantly blue relative to the bulk population of passive BCGs at a similar redshift. Since 20/22 of these sources are identified as BL~Lac objects this is as expected, given that one of the characteristics of BL~Lacs is a strong blue continuum in their spectra \citep{Massaro+12}. This further supports the validity of the AGN identification.

\subsection{Previously Identified AGN in BCGs}\label{sect:rediscover}

In our analysis we have independently rediscovered a number of sources which have already been identified as an AGN hosted by BCG. These cover a broad range in terms of the relative AGN-to-cluster X-ray contributions, providing valuable context for the candidate systems. Where given below, the relative contribution of AGN and cluster emission are estimated from summing the relative photon counts, in \textit{XMM-Newton} and \textit{Chandra} imaging, which can roughly be attributed to point-source-like emission and the more extended emission (similar to the approach outlined in \citealt{Russell+13}). As such, they are only approximate values and do not take into consideration X-ray variability or the spectral energy distribution of the components.

\indent \textbf{1H1821+644:}
This highly luminous broad-line quasar resides in the BCG of a high mass, strong cool core cluster. Interestingly, the ICM around the quasar has a significantly lower entropy and temperature and steeper gradients than comparable clusters \citep{Walker+14}. Within 80 kpc the gradients and profiles are consistent with those of cooling flows, suggesting the AGN is failing to heat ICM and offset radiative cooling as expected. From the \textit{Chandra} analysis \cite{Russell+10}, we know the quasar dominates the central region, but that the cluster is also a significant contributor of X-rays. The relative X-ray contributions are estimated as roughly one part cluster to two parts quasar. The position of this AGN in \Lx--Richness space (Fig. \ref{fig:richness-Lx}) is as expected for an X-ray source dominated by the AGN emission, characterised by a relatively low richness for a high \Lx. Similarly the high \Lx relative to the BCG luminosity is as expected for an AGN dominated system. The extreme optical and MIR luminosity, and MIR colours, indicate the dominant nature of the quasar at these wavelengths. 

\indent\textbf{H0414+009:}
For a long time H0414+009 has been known to be a BL~Lac object hosted by a BCG (\citealt{McHardy+92,Falamo+93}). Archival \textit{XMM-Newton} and \textit{Chandra} reveal clear point source domination. This is fully consistent with the significant offset in both richness and BCG luminosity from the comparison cluster sample. 
  
\indent \textbf{A2055:}
This is an example of a BL~Lac object in the BCG of a known Abell cluster. From an archival \textit{XMM-Newton} observation we estimate the relative cluster--AGN contribution as approximately 1:4. This is perhaps surprising as we expected a more comparable mix of two from its richness--\Lx position. However, if we consider that the estimated cluster contribution is $\sim0.5\times10^{44}\erg$, its position in richness--\Lx space is still comfortably comparable to the cluster sample.

\indent \textbf{RXJ0909.9+3106:}
This appears to be a case of a BL~Lac in the BCG of a group of galaxies at a redshift of 0.272. An archival \chandra observation shows the presence of extended cluster emission, confirming the presence of a cluster, but the X-ray emission is clearly AGN dominated (by factor of $\sim20$). This is consistent with our predictions based on the richness--\Lx position and the optical/MIR luminosities. Note, a drop in \Lx by a factor of 20 would place it comfortably amongst the comparison cluster richness--\Lx distribution. 

\indent \textbf{RXJ1745+3951:}
This BL Lac object has a \rosat HRI observation which shows extended cluster emission. Although the core could not be resolved the relative contributions of the BL~Lac and the cluster were estimated by \cite{Gliozzi+99} to be roughly equal. An optical analysis revealed the presence of a gravitational arc from a strongly lensed $z=1.06$ galaxy (\citealt{Nilsson+99}) confirming a high cluster mass, later estimated to be $4 \times 10^{14}\mathrm{M}_{\sun}$ \citep{Lietzen+08}. The richness and BCG luminosities are fully consistent with other clusters at similar \Lx, as expected if this roughly equal mix is genuine. 
\begin{table*}
 \centering
 \caption{Systems which are likely AGN dominated (red stars in figures).  (a) X-ray luminosity in units of $10^{44}~\erg$, (b) red sequence richness.}
\begin{tabular}{lcccrcll}
  \hline
  \hline
\centering Cluster/AGN ID & BCG R.A. & BCG DEC & Redshift & \Lx & Rich. & AGN Class & Comments\\
\centering &  (J2000) & (J2000) & & (a)&(b)\\
\hline 
 1H0414+00       & 04:16:54.5 & $+01$:04:56  & 0.287 & 118.00 & 11 & BL~Lac & \textit{Chandra}/\textit{XMM}: point source dominated  \\ 
 RXJ0828.2+4154  & 08:28:14.2 & $+41$:53:52  & 0.226 & 2.46 & 19 & BL~Lac & \textit{Chandra}: point source dom. (Donahue priv. comm.)\\ 
RXJ0909.9+3106 & 09:09:53.3 & $+31$:06:03  & 0.272 & 12.00 & 23 & BL~Lac & \textit{Chandra}: extended, but AGN dominated \\ 
 RXJ1117.1+2014  & 11:17:06.3 & $+20$:14:07  & 0.138 & 29.40 & 20 & BL~Lac & NORAS ``AGN-Cluster'' ID  \citep{Bohringer+00}\\ 
 RXJ1140.4+1528  & 11:40:23.5 & $+15$:28:09  & 0.244 & 13.30 & 19 & BL~Lac & \\
 RXJ1442.8+1200  & 14:42:48.1 & $+12$:00:40  & 0.163 & 8.72 & 17 & BL~Lac & Dumbbell; \textit{Chandra}: point source dominated \\ 
 RXJ1617.1+4106  & 16:17:06.3 & $+41$:06:47  & 0.267 & 4.86 & 12 & BL~Lac\\ 
1H1821+644     & 18:21:54.4 & $+64$:20:09  & 0.297 & 57.80 & 41 & Broad-line& \textit{Chandra}: extended emission, but quasar dominates \\ 
 RXJ2113.9+1330  & 21:13:53.7 & $+13$:30:17  & 0.307 & 4.70 & 10 & BL~Lac\\ 

 RXJ2150.2-1410  & 21:50:15.5 & $-14$:10:49  & 0.229 & 17.50 & 17 & BL~Lac & \\ 

\hline
\label{Table:redstars}
\end{tabular}
\end{table*}
\begin{table*}
 \centering
 \caption{Systems which are likely group or poor cluster scales (yellow stars in figures). (a) X-ray luminosity in units of $10^{44}~\erg$, (b) red sequence richness.}
\begin{tabular}{lcccrcll}
  \hline
  \hline
\centering Cluster/AGN ID &  BCG R.A. & BCG DEC & Redshift & \Lx & Rich. & AGN Class & Comments\\
\centering & (J2000) & (J2000) & & (a) & (b) &\\
\hline
 RXJ0039.1-2220  & 00:39:08.2 & $-22$:20:01  & 0.065 & 0.22 & 23 & Broad-line & \\
 RXJ0110.1+4150  & 01:10:04.8 & $+41$:49:50  & 0.096 & 0.26 & 14 & BL~Lac & \\ 
 RXJ0656.2+4237  & 06:56:10.7 & $+42$:37:03  & 0.059 & 0.22  & 31 & BL~Lac & \\
 RXJ1053.7+4930  & 10:53:44.1 & $+49$:29:56  & 0.140 & 0.70 & 17 & BL~Lac & \\
 RXJ2336.9-2326  & 23:36:53.8 & $-23$:26:26  & 0.160 & 0.31 & 18 & BL~Lac \\ 

\hline
\label{Table:othercandidates}
\end{tabular}
\end{table*}

\begin{table*}
 \centering
 \caption{Best candidate systems of AGN in BCG (blue stars in figures). (a) X-ray luminosity in units of $10^{44}~\erg$, (b) red sequence richness.}
\begin{tabular}{lcccrcll}
  \hline
  \hline
\centering Cluster/AGN ID & BCG R.A. & BCG DEC & Redshift & \Lx & Rich. & AGN Class & Comments/X-rays Obs.\\
\centering &  (J2000) & (J2000) && (a) &(b)& \\
\hline 
RXJ0014.3+0854 & 00:14:19.7 & $+08$:54:02  & 0.163 & 0.85 & 36 & BL~Lac \\ 
RXJ0056.3-0936 & 00:56:20.1 & $-09$:36:32  & 0.103 & 1.86 & 23 & BL~Lac & BCG is a dumbbell galaxy\\ 
RXJ1013.6-1351 & 10:13:35.2 & $-13$:51:29  & 0.152 & 1.53 & 44 & BL~Lac &\\ 
RXJ1516.7+2918 & 15:16:41.6 & $+29$:18:09  & 0.130 & 0.84 & 23 & BL~Lac &\\ 
A2055          & 15:18:45.7 & $+06$:13:56  & 0.102 & 2.49 & 58 & BL~Lac & Cluster confirmed: \textit{XMM-Newton} \\ 
RXJ1745.6+3951 & 17:45:37.7 & $+39$:51:31  & 0.267 & 2.73 & 38 & BL~Lac & Cluster confirmed: \rosat HRI\\ 
RXJ2353.4-1257 & 23:53:25.1 & $-12$:57:01  & 0.240 & 2.01 & 30 &  BL~Lac & \\ 
\hline
\label{Table:bestcandidates}
\end{tabular}
\end{table*}

\subsection{Clusters Previously Incorrectly Identified as AGN}

There are a number of systems in which the identification of
the majority of the X-ray emission has alternated between
cluster and AGN. We give two illustrative examples.

\indent \textbf{Zw2089:}
This is a LoCuSS cluster that shows relatively broad lines 
in its optical spectrum, but the point
source does not contribute more than 5\% to the total
cluster flux \citep{Russell+13}. 

\indent \textbf{RXCJ0132.6-0804:}
\cite{Dutson+13} find that this cluster may contain
a {\it Fermi} source in its BCG, however an archival \chandra
observation finds the majority of the emission
is from the extended emission. 

Therefore, it is important to be aware that many systems
are a combination of both cluster and AGN emission in 
X-ray data. This is illustrated in the compilation of BL~Lac
samples, such as the Roma BZCAT sample, where the radio and
X-ray properties of clusters can be misinterpreted as
being from a BL~Lac.

\subsubsection{The Roma BZCAT catalogue}\label{sect:roma}

\begin{table*}
 \centering
 \caption{BZCAT BL~Lac candidates that are likely to be clusters. The
Cluster IDs marked with an asterisk have archival X-ray pointed
data that in all cases shows the emission to be dominated
by extended emission.}
\begin{tabular}{llcl}
  \hline
  \hline
\centering BZCAT ID & Cluster ID & Redshift & Reference\\
\centering &  & & \\
\hline
5BZGJ0006+1051 & Z15              & 0.1676 &  eBCS \cite{Ebeling+00} \\
5BZGJ0012-1628 & A11              & 0.1510 &  RBSC \cite{Bauer+00} \\
5BZGJ0014+0854 & MS0011.7+0837    & 0.1633 &  EMSS \cite{Stocke+91} \\
5BZGJ0027+2607 & MACSJ0027.4+2607* & 0.3645 &  MACS \cite{Ebeling+10} \\
5BZGJ0056-0936 & RXJ0056.3-0936    & 0.1031 &  This paper \\
5BZGJ0123+4216 & NVSS cluster     & 0.1856 &  NVSS \cite{Bauer+00} \\
5BZGJ0153-0118 & NVSS cluster     & 0.2458 &  NVSS \cite{Bauer+00} \\
5BZGJ0425-0833 & EXO0422-086      & 0.0390 &  REFLEX \cite{Bohringer+04} \\
5BZGJ0439+0520 & RXCJ0439.0+0520*  & 0.2080 &  BCS \cite{Ebeling+98}\\
5BZGJ0656+4237 & RXJ0656.1+4237   & 0.0590 &   This paper\\
5BZGJ0737+5941 & UGC3927          & 0.0410 &   \\
5BZGJ0737+3517 & A590             & 0.2104 &   \\
5BZGJ0751+1730 & Z1432            & 0.1866 &   eBCS \cite{Ebeling+00} \\
5BZGJ0753+2921 & A602             & 0.0621 &    BCS \cite{Ebeling+98} \\
5BZGJ0834+5534 & 4C+55.16*        & 0.2412 & \cite{Iwasawa+99} \\
5BZGJ0856+5418 &                  & 0.2594 & RBSC \cite{Bauer+00}\\
5BZGJ0927+5327 & Z2379*           & 0.2011 & eBCS \cite{Ebeling+00}\\
5BZGJ0944-1347 & NVSS cluster     & 0.1782 & RBSC \cite{Bauer+00} \\
5BZGJ1108-0149 & NVSS cluster     & 0.1056 & RBSC \cite{Bauer+00} \\
5BZGJ1119+0900 & RedMAPPER        & 0.3315 &  Chandra obs \\
5BZGJ1350+0940 & RXCJ1350.3+0940* & 0.1325 &  RBSC \cite{Hogan+15b} \\
5BZGJ1356-3421 & PKS1353-341*     & 0.2227 &  \cite{VC+00} \\
5BZGJ1407-2701 & A3581*           & 0.0218 &  REFLEX \cite{Bohringer+04}\\
5BZGJ1413+4339 & A1885*           & 0.0893 &  BCS \cite{Ebeling+98} \\
5BZGJ1445+0039 &                  & 0.3062 &  \\
5BZGJ1459-1810 & S780*            & 0.2360 &  REFLEX \cite{Bohringer+04}\\
5BZGJ1504-0248 & RXCJ1504.1-0248* & 0.2169 &  REFLEX \cite{Bohringer+04}\\
5BZGJ1516+2918 & RXJ1516.7+2918   & 0.1298 &  This paper \\
5BZGJ1532+3020 & MACSJ1532.8+3021* & 0.3621 &  BCS \cite{Ebeling+98}\\
5BZGJ1603+1554 & RXCJ1603.6+1553*  & 0.1097 &  RBSC \cite{Hogan+15a}\\
5BZGJ1715+5724 & NGC6338*          & 0.0281 &  BCS \cite{Ebeling+98}  \\
5BZGJ1717+2931 & RBS1634          & 0.2782 &  RBSC \cite{Bauer+00} \\
5BZGJ1717+4227 & Zw8193*          & 0.1829 &  eBCS\cite{Ebeling+00} \\
5BZGJ1727+5510 & A2270*           & 0.2473 &  RBSC \cite{Hogan+15a}\\
5BZGJ1745+3951 & RGB 1745+398*    & 0.2670 &  \cite{Gliozzi+99} \\
5BZGJ1804+0042 & CIZA1804.1+0042  & 0.0700 &  CIZA \cite{Ebeling+00}\\
5BZGJ2003-0856 & NVSS cluster     & 0.0572 &  RBSC \cite{Bauer+00} \\
5BZGJ2140-2339 & MS2137.3-2353*   & 0.3130 &  EMSS \cite{Stocke+91}\\
5BZGJ2147-1019 & REFLEX cluster     & 0.0797 &  REFLEX \cite{Bohringer+04} \\
5BZGJ2320+4146 & CIZA2320.2+4146  & 0.1520 &  CIZA \cite{Ebeling+00}\\
5BZGJ2341+0018 & NORAS cluster    & 0.2767 &  NORAS \cite{Bohringer+04}\\
\hline
\label{Table:BZCAT}
\end{tabular}
\end{table*}

Table \ref{Table:BZCAT} lists the 41 ROSAT BSC sources which are listed
in \cite{Massaro+15}
as being BL~Lac objects, but in which we
are confident the cluster emission dominates. These results are based on a combination of this analysis, archival X-ray data, and the literature. 
The large majority are listed in \cite{Massaro+15} as ``BL~Lac-galaxy dominated'', meaning their SEDs suggest dominance of
the host galaxy over the nuclear emission. We note their optical spectra show strong emission
lines in almost all cases (e.g. 4C+55.16/5BZGJ0834+5534, \citealt{Iwasawa+99}. 
Clusters with pointed X-ray observations are noted in
the table and in all cases the extended emission
dominates. While the selection method used to
create the BZCAT sample is efficient in selecting
BL~Lacs, there is a significant number of sources
that need to be filtered out to avoid contamination.
Using the fraction of polarisation in the radio, as used by
\cite{Edge+03}, or the detection of significant
extent in the RASS emission \citep{Voges+99}, would
have excluded the majority of the sources listed
in Table \ref{Table:BZCAT}, without significantly affecting the
selection of BL~Lacs.

\subsection{Lessons for the Future}

The pending launch of {\sl eROSITA} and subsequent X-ray survey will dramatically
increase the number of clusters and AGN detected in hard X-rays. The importance
of having a clear context in which to place the more ambiguous sources, where
there is evidence of both a rich cluster {\it and} an AGN, means that we
should consider what this study tells us.

Firstly, the number of X-ray selected AGN that reside at the cores
of rich clusters is relatively low. There are not many BCGs that
are accreting close to their Eddington rate and lost to 
our X-ray selection. This is to be expected from the likely
duty cycle of AGN and we can set a more stringent upper
limit to this by noting that very few of the systems 
identified here are broad-lined AGN. Instead the large majority are BL~Lacs,
where orientation effects dominate. Therefore, given
our parent sample of X-ray clusters is nearly 1000
in total (\citealt{Green+16}), the duty cycle for 
high accretion rate events is well below 1\%,
with just two clearly identifiable broad-line
AGN in our sample. 

We can also place a crude
limit on the angular size of the relativistic beaming
cone for jets in BCGs. Assuming the $\sim$7 BL~Lacs
are drawn randomly in orientation from all BCGs, in the total of 949
$z<0.4$ RASS clusters covered by the PS1 footprint,
then only 0.74\% of BCGs are beamed 
towards us. So the opening angle is 7.0$^\circ$.
If we exclude BCGs with low
radio power, (approximately two thirds of the sample)
using the radio luminosity function of BCGs in \cite{Hogan+15a},
then the fraction of beamed objects increases to approximately 2.2\%
and hence the implied opening angle is 12.1$^\circ$.
This is consistent with limits from superluminal motion
studies in BL Lacs and beamed flat-spectrum radio galaxies, 
which suggest angles of 2--5$^\circ$ 
for Lorentz factors less than 10,
as should be the case in these sources (\citealt{Jorstad+05}).
Clearly a significantly better constraint can be derived
in the near future with the combination of
{\it eROSITA} and new radio surveys such as VLASS,
EMU and WODAN.

Although outside of the immediate scope of this work, 
a further subtlety to consider, in identifying ambiguous sources 
and understanding the 
feedback of AGN in cluster cores, is AGN variability.
Variability has the potential to tip the balance of 
emission between cluster and AGN contributions at the time of 
some given observation. An illustrative example is NGC1275 in the
Perseus cluster (\citealt{ODea+84,Dutson+14,Fabian+15}), which
has exhibited a variation of at least
one order of magnitude on a timescale
of a few decades.

\section{Summary and Conclusions}\label{sect:conclude}
We use the PS1 $3\pi$ survey to look for signs of a rich cluster around the full \rosat All Sky Survey AGN population in the PS1 $3\pi$ footprint, constituting some 3058 broad-line AGN and 412 BL~Lac objects. Of these, we find 22 AGN with signs of a significant overdensity in red galaxies, suggestive of a possible cluster red sequence, and which are hosted by a galaxy visually consistent with being a BCG. We compare the properties of these candidate systems with those of a comparison sample of the $\sim1000$ confirmed clusters also drawn from the \rosat All Sky Survey.

\indent We identify seven best candidates, contingent on having a red sequence richness that is most comparable to the confirmed massive clusters at similar X-ray luminosities. Amongst our 22 candidate systems, five have been previously confirmed to be a cluster with a BCG in its central galaxy, with pointed X-ray observations available, strongly validating our technique, and providing valuable context for our candidate systems.  

\indent We use the colour of the red sequence as a tracer of redshift to investigate cluster membership of the AGN. We find all our candidate systems have a red sequence colour which is similar to that of the confirmed clusters at similar redshift, indicative that the AGN and red sequence galaxies share a common redshift. 

\indent The optical and MIR luminosities of the AGN host galaxies are found to be generally consistent with those of the BCGs of the comparison clusters, at similar X-ray luminosities. This supports our visual identification of the AGN host galaxies as BCGs.

\indent We also check the validity of the initial AGN identification by looking at the MIR and optical colours of our AGN. In the MIR we find all candidate systems show an excess MIR emission with respect to the bulk passively evolving population of comparison BCGs, suggestive of activity. In the optical we find the majority of sources are significantly blue in \textit{g}--\textit{r}, with respect to the passively evolving BCGs, consistent with their identification as BL~Lacs.

\indent We emphasise the need to consider the ambiguity that exists in the identification of clusters and AGN in X-ray data, and stress it is not necessarily a case of one of the other. This very simple approach of checking for optical overdensities, around X-ray selected AGN, can be utilised to correct for selection effects against the detection of strong AGN in the cores of clusters. This ultimately will provide better statistics relating to both the prevalence of, and energetics involved with, AGN in BCGs  - increasing our understanding of the role of AGN feedback in this unique environment. This consideration will be particularly important in the future with the next generation of X-rays survey telescopes, such as \textit{eROSITA}, set to uncover vast numbers of new X-ray sources; including clusters, AGN \textit{and} the two combined.

\section*{Acknowledgements}
We thank the referee for their valuable comments and suggestions to the improvement of this manuscript. TSG received financial support from the Science and Technology Facilities Council (STFC) grant ST/K501979/1. ACE acknowledges support from STFC grant ST/I001573/1. 

This research has made use of the NASA/IPAC Extragalactic Database (NED) which is operated by the Jet Propulsion Laboratory, California Institute of Technology, under contract with the National Aeronautics and Space Administration.
 
The Pan-STARRS1 Surveys (PS1) have been made possible through contributions by the Institute for Astronomy, the University of Hawaii, the Pan-STARRS Project Office, the Max-Planck Society and its participating institutes, the Max Planck Institute for Astronomy, Heidelberg and the Max Planck Institute for Extraterrestrial Physics, Garching, The Johns Hopkins University, Durham University, the University of Edinburgh, the Queen's University Belfast, the Harvard-Smithsonian Center for Astrophysics, the Las Cumbres Observatory Global Telescope Network Incorporated, the National Central University of Taiwan, the Space Telescope Science Institute, and the National Aeronautics and Space Administration under Grant No. NNX08AR22G issued through the Planetary Science Division of the NASA Science Mission Directorate, the National Science Foundation Grant No. AST-1238877, the University of Maryland, Eotvos Lorand University (ELTE), and the Los Alamos National Laboratory.

\indent AllWISE makes use of data from WISE, which is a joint project of the University of California, Los Angeles, and the Jet Propulsion Laboratory/California Institute of Technology, and NEOWISE, which is a project of the Jet Propulsion Laboratory/California Institute of Technology. WISE and NEOWISE are funded by the National Aeronautics and Space Administration.
  



\bibliographystyle{mnras}
\bibliography{paper} 



\appendix
\section{Colour--Magnitude Diagrams and Colour Composite Images}
Here we show the $g$--$r$ colour-magnitude diagram, as well as $gri$ colour images at cluster and BCG scale, for the `best' candidates and the previously identified systems.

\begin{figure*}
\begin{minipage}{\textwidth}
{\includegraphics[width = 0.4\textwidth]{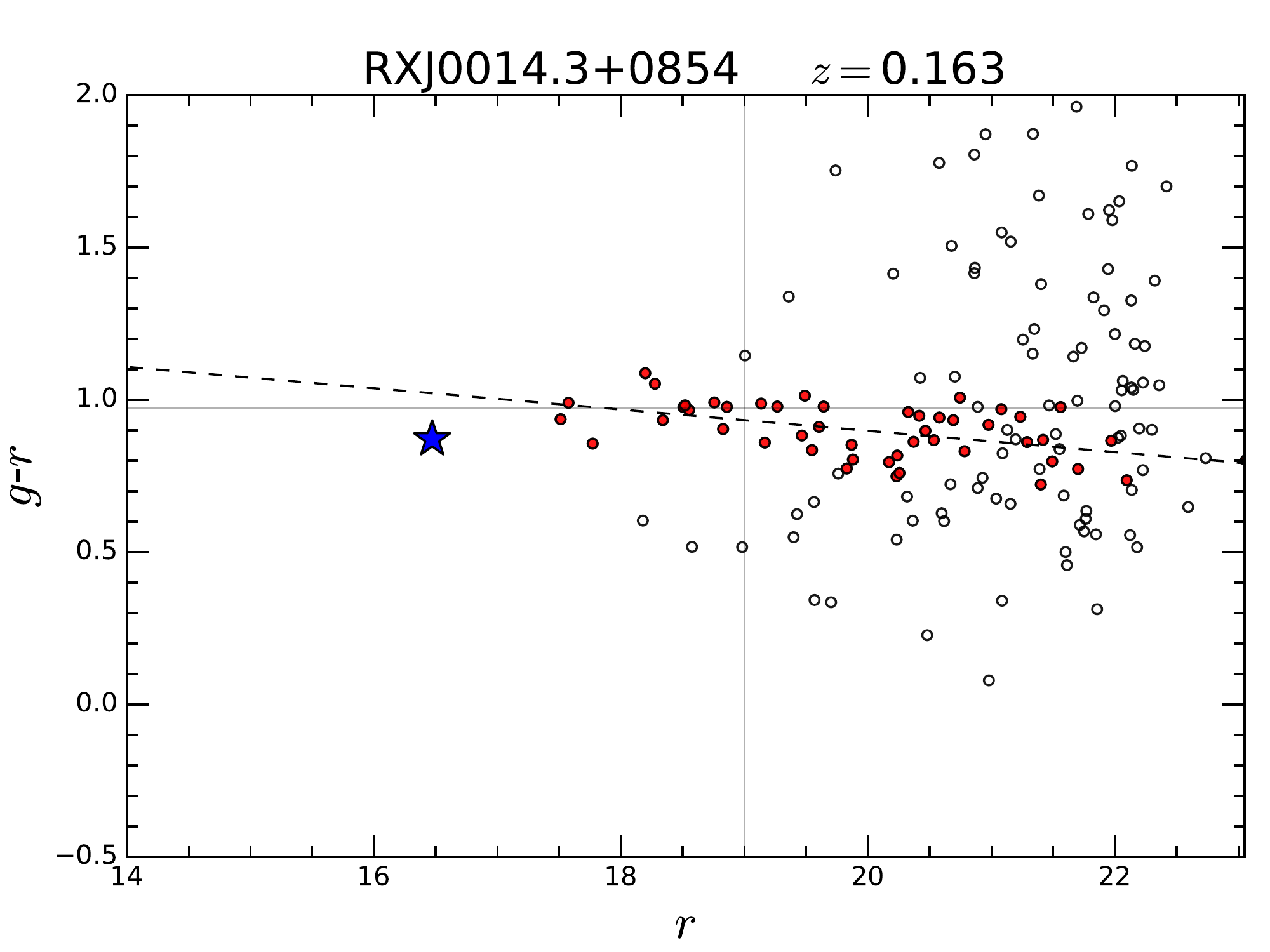}}
{\includegraphics[width = 0.6\textwidth]{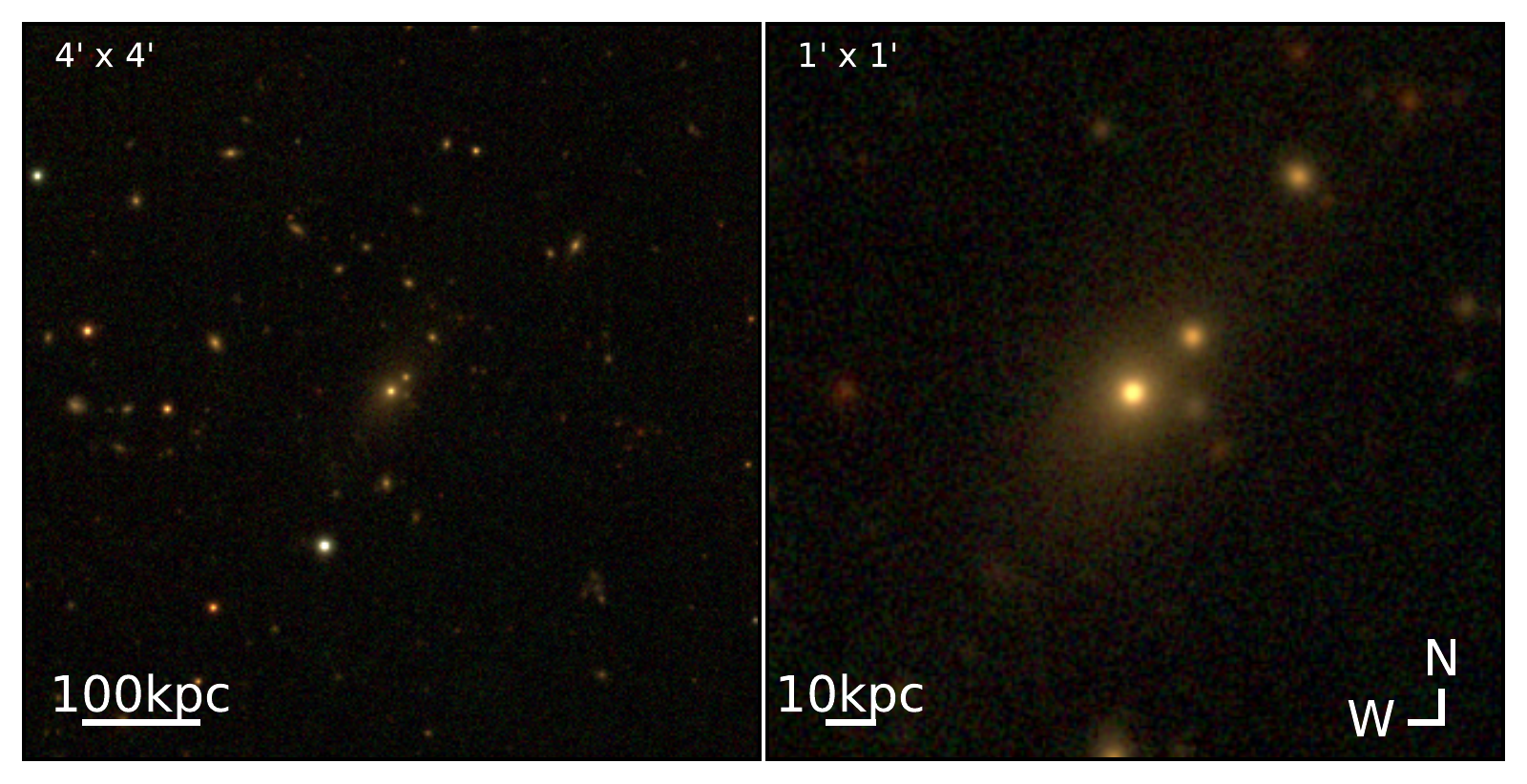}}
{\includegraphics[width = 0.4\textwidth]{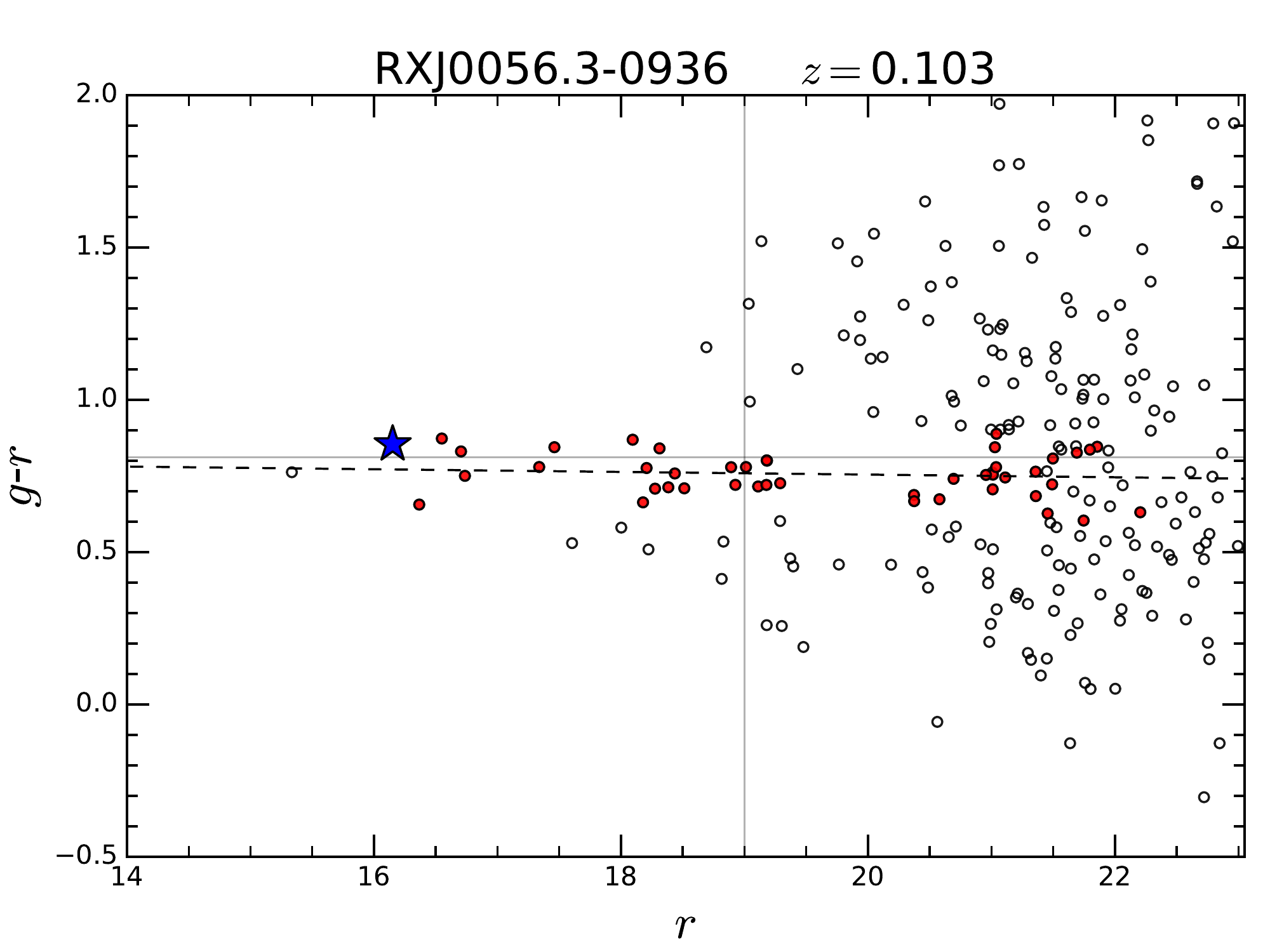}}
{\includegraphics[width = 0.6\textwidth]{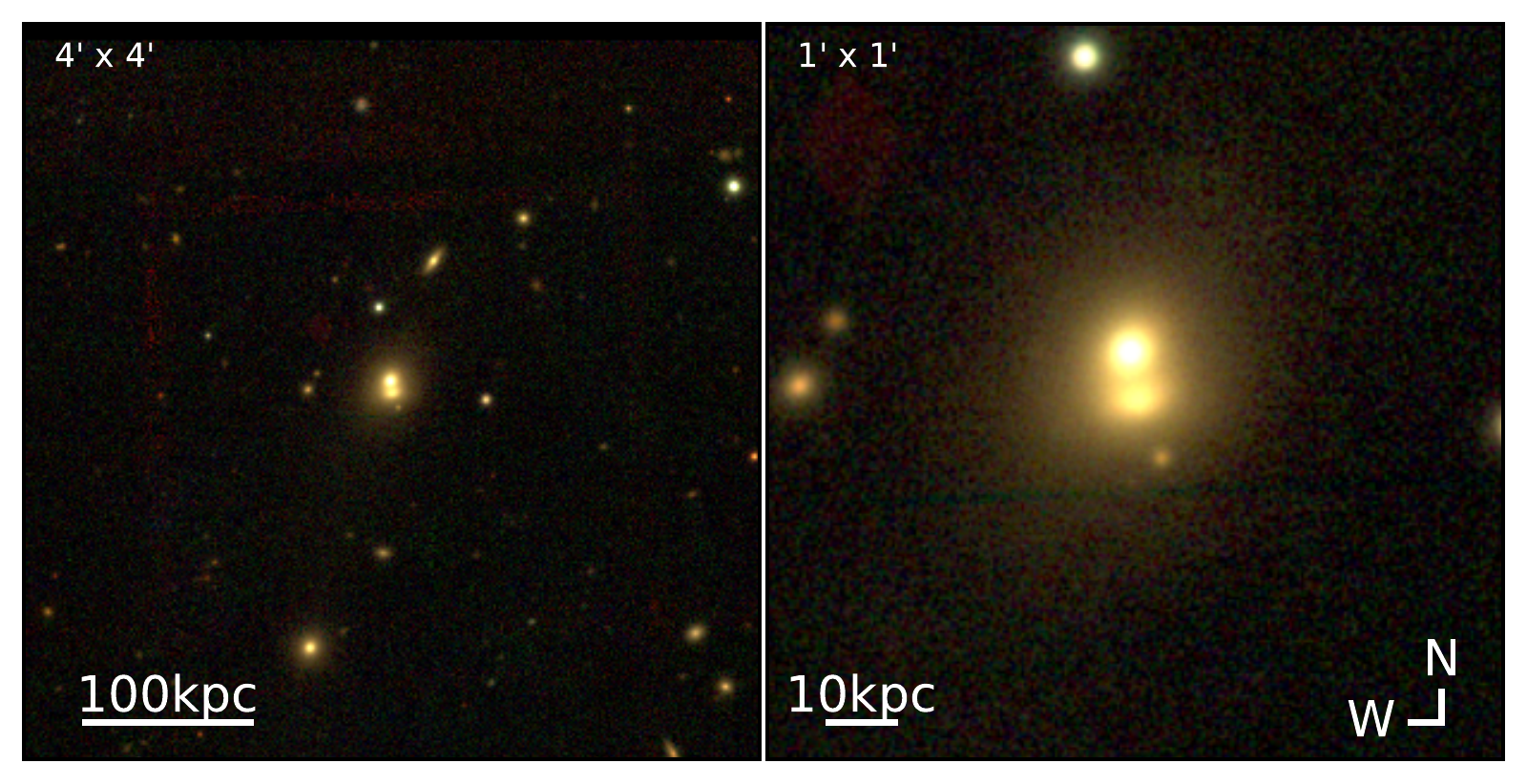}}
{\includegraphics[width = 0.4\textwidth]{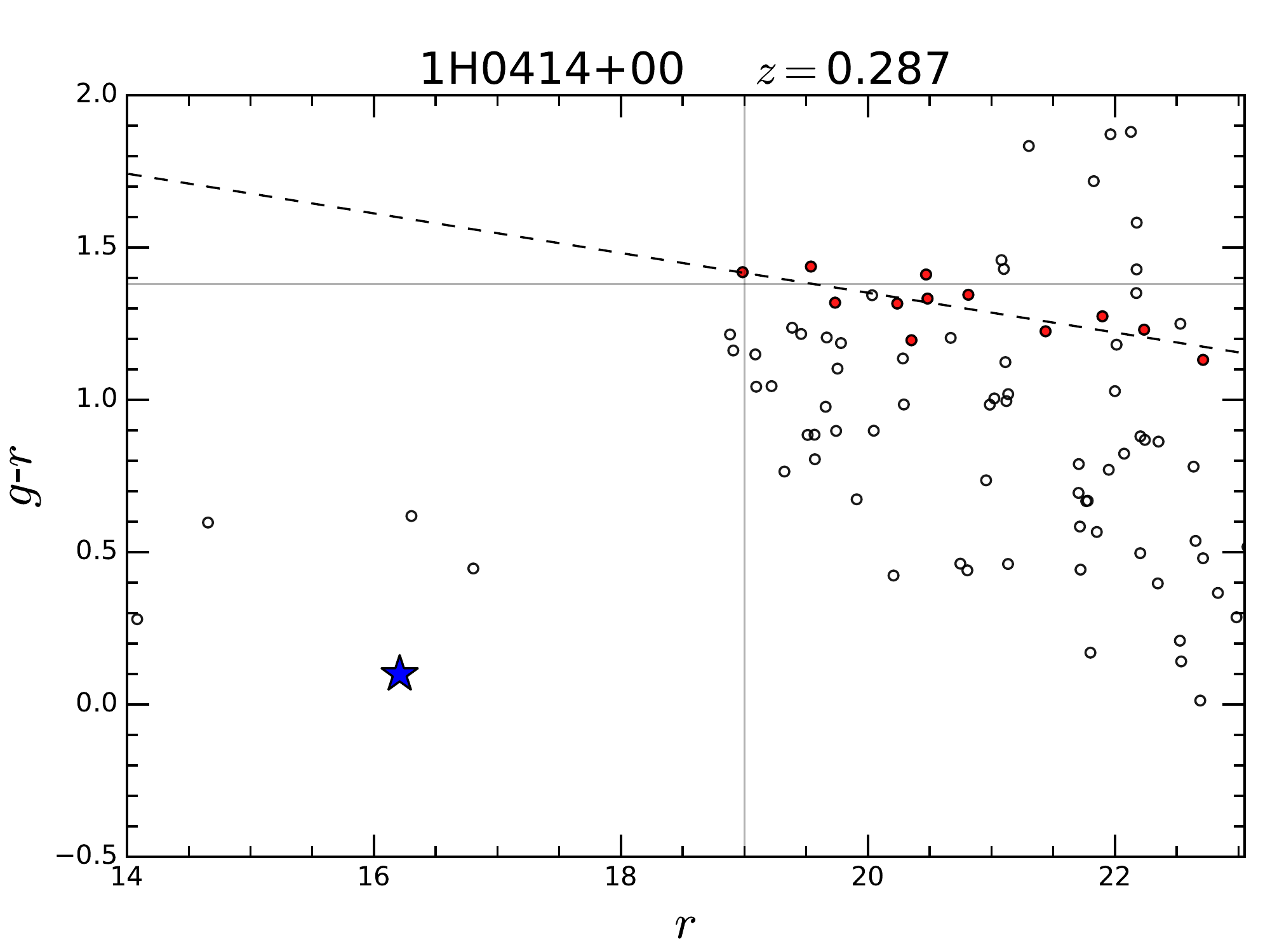}}
{\includegraphics[width = 0.6\textwidth]{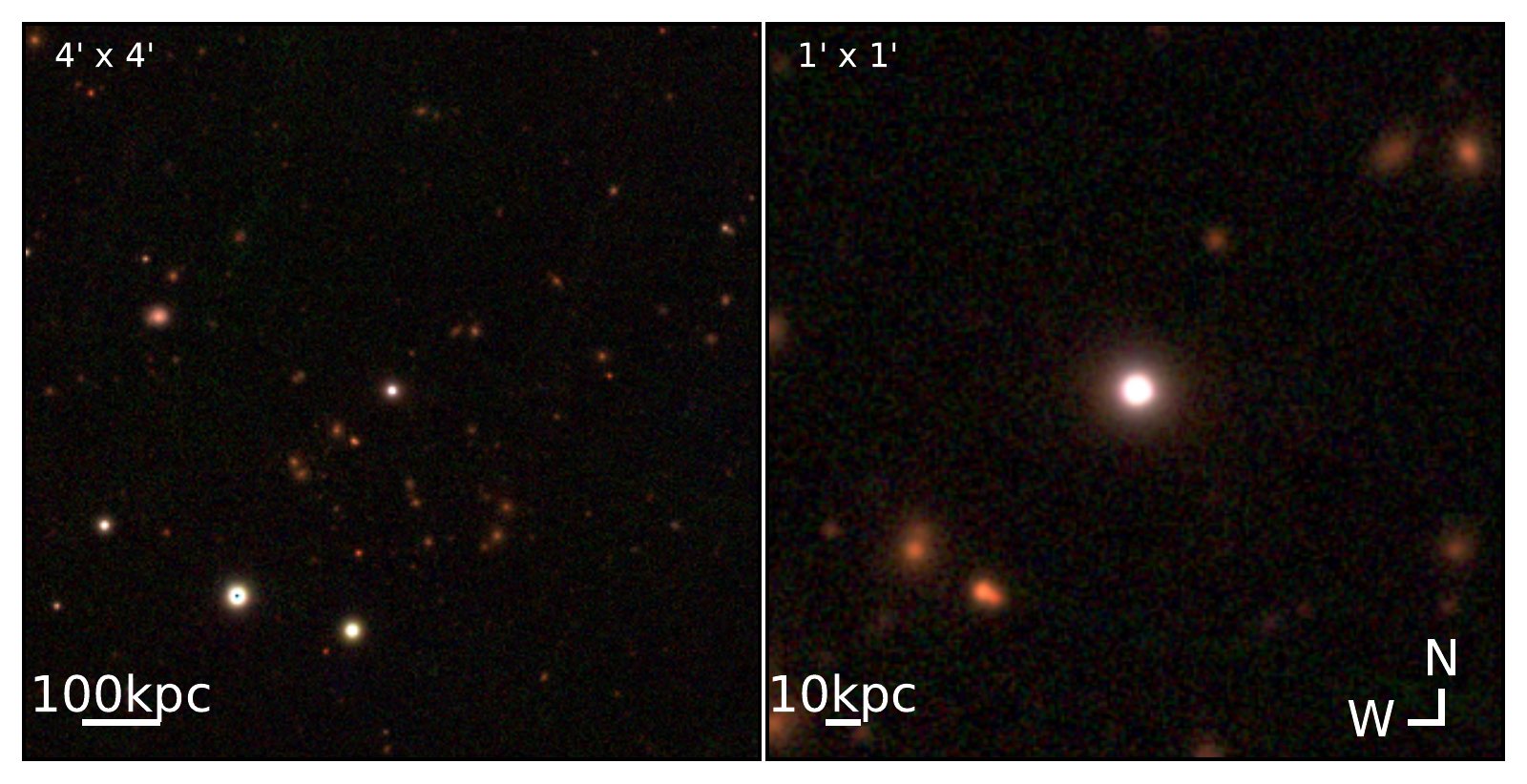}}
{\includegraphics[width = 0.4\textwidth]{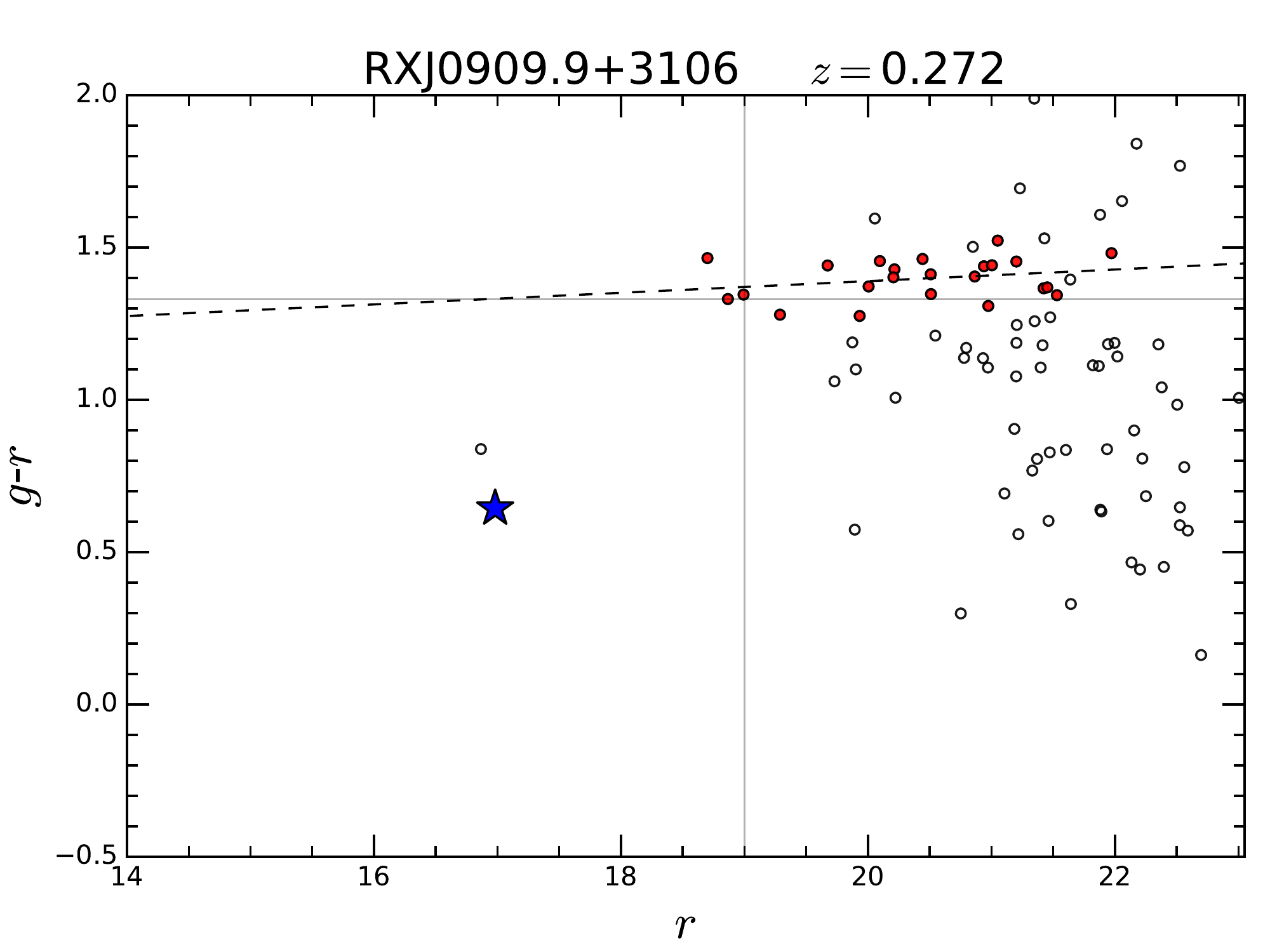}}
{\includegraphics[width = 0.6\textwidth]{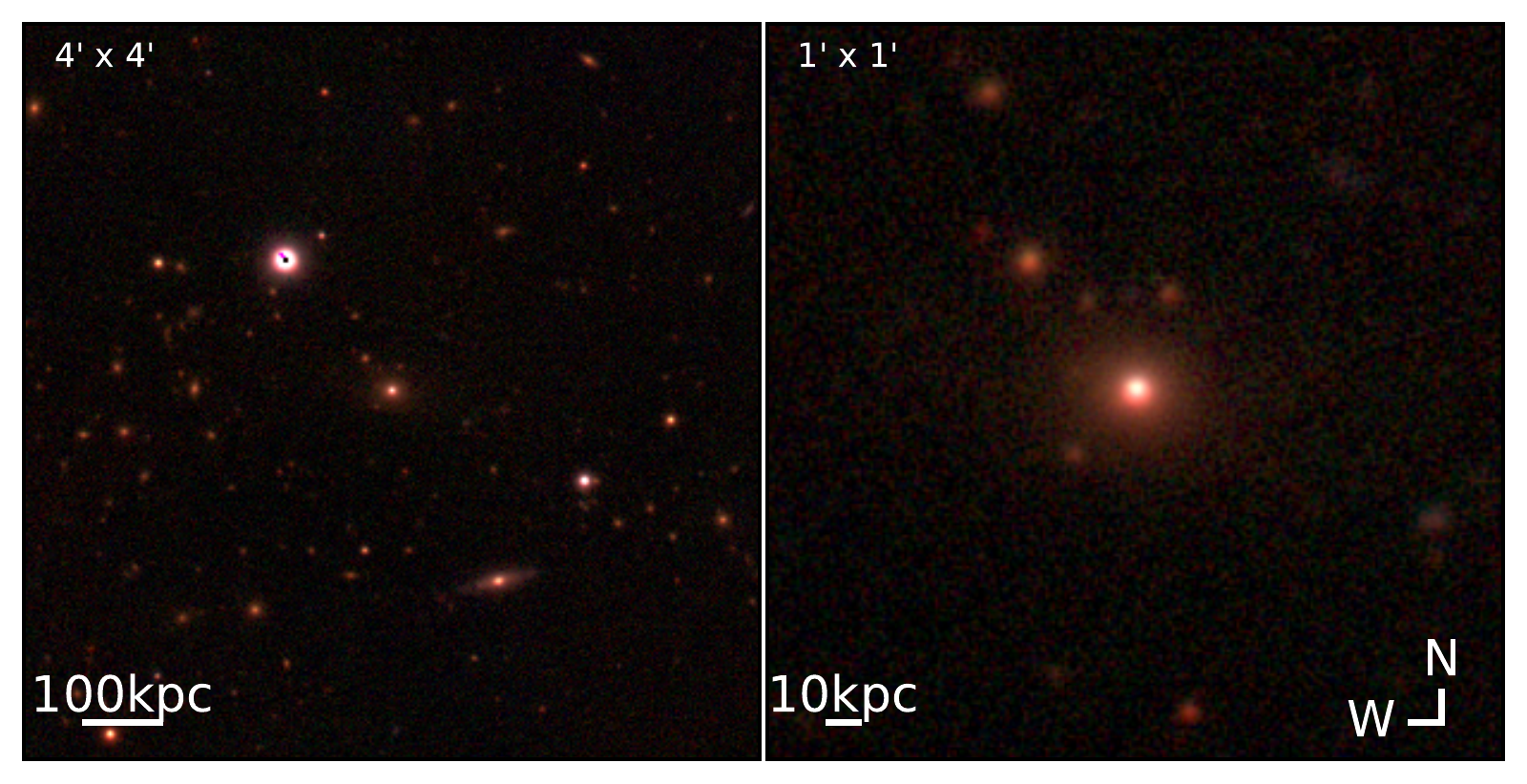}}
\end{minipage}
    \caption{Left: The PS1 \gr colour-magnitude diagram for sources within a radius of $0.5\,\Mpc$ around the AGN position. The filled (red) circles indicate galaxies selected to be on the cluster red sequence and the dashed line shows a best fitting linear regression line to these points. The (blue) stars indicate the AGN host galaxy, believed to be the BCG. And, the intersection of the solid grey lines indicates the predicted red sequence colour, at 19th magnitude, given its redshift (as estimated from our comparison cluster sample -- i.e. Fig. \ref{fig:RScol-z}).
Middle: \textit{gri} colour image in a $4\times4$ arcmin box, centred on the BCG. Right: \textit{gri} colour image in a $1\times1$ arcmin box, centred on the AGN/BCG.}
    \label{fig:CMR}
\end{figure*}

\begin{figure*}
\begin{minipage}{\textwidth}
{\includegraphics[width = 0.4\textwidth]{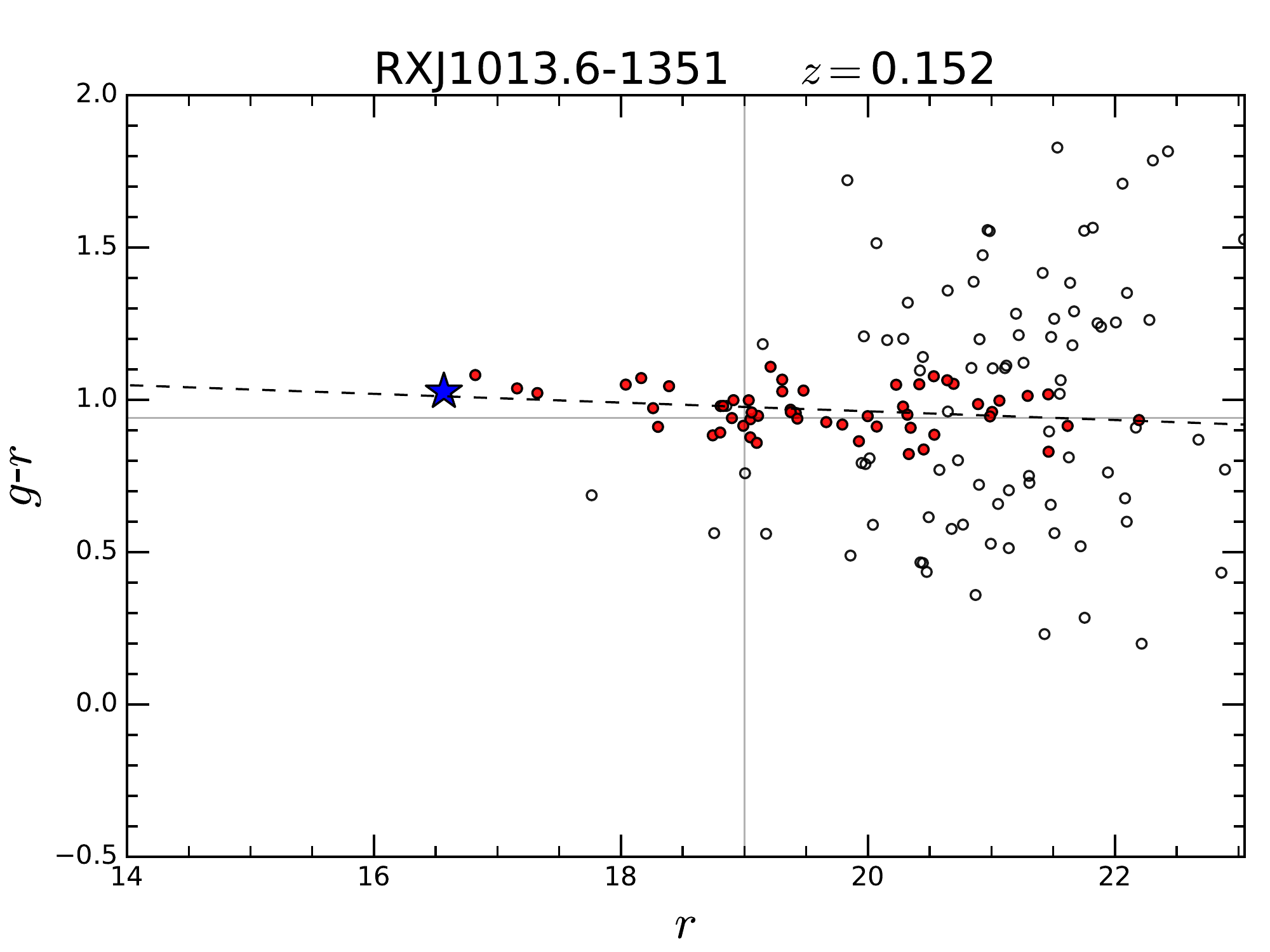}}
{\includegraphics[width = 0.6\textwidth]{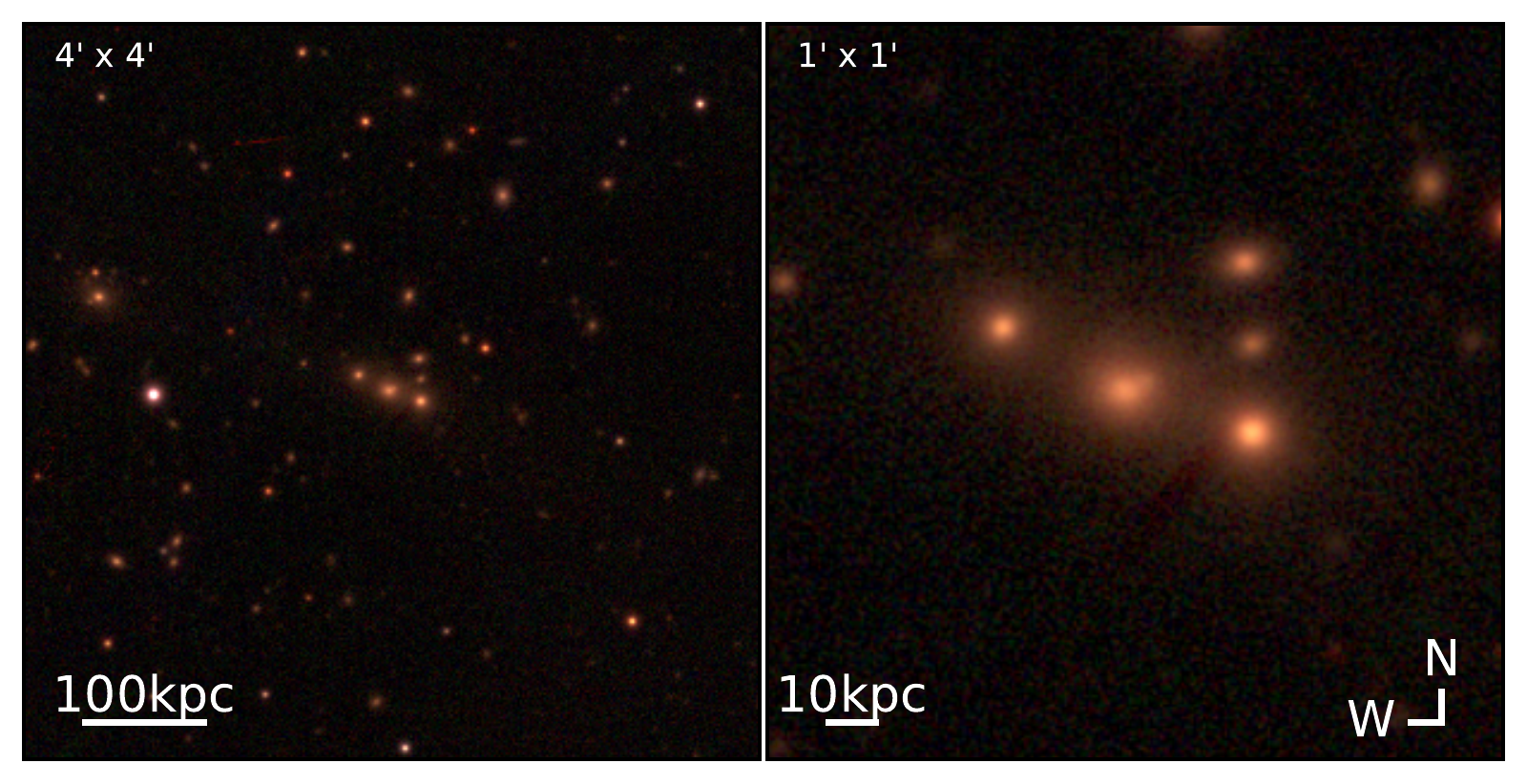}}
{\includegraphics[width = 0.4\textwidth]{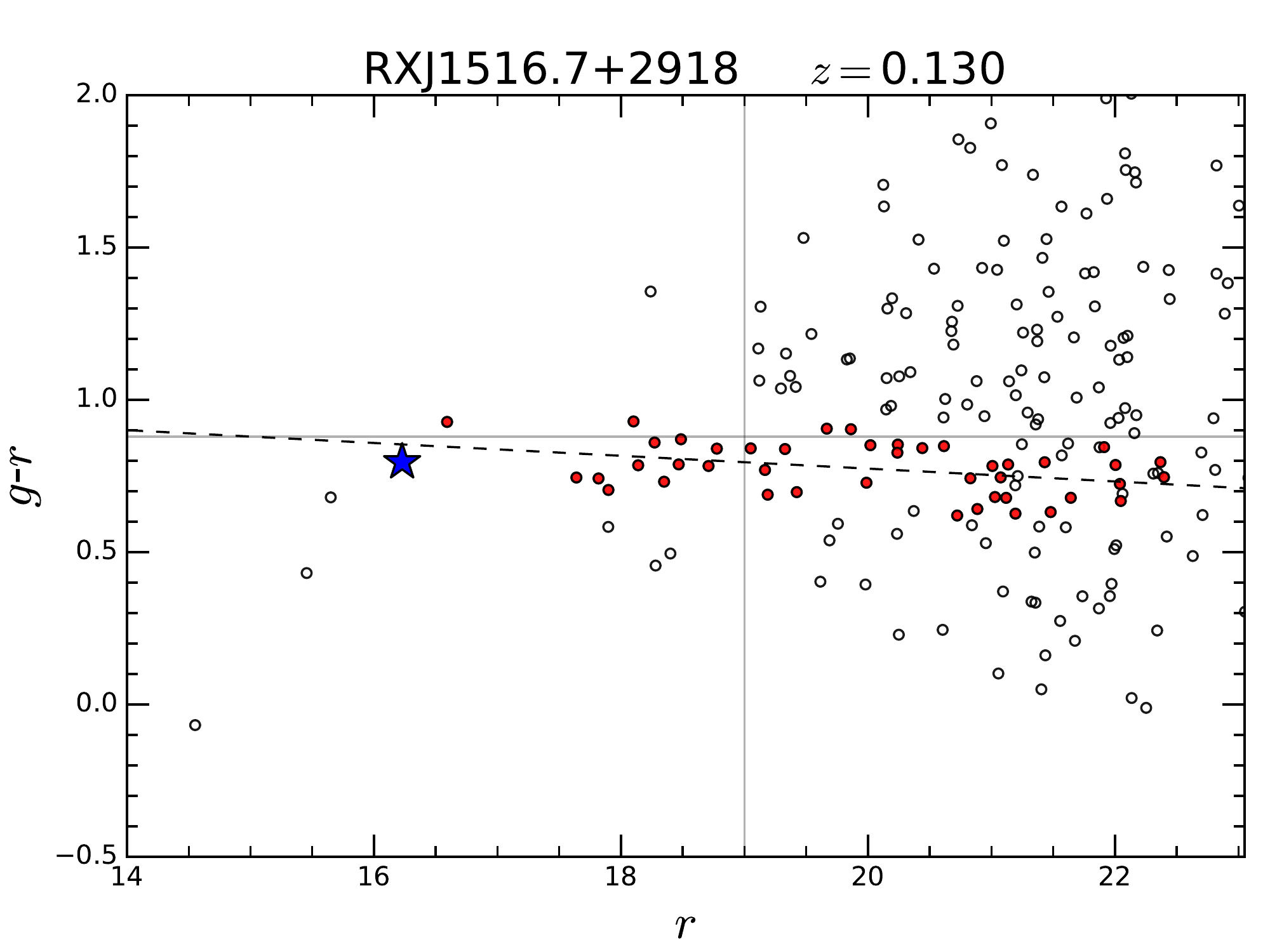}}
{\includegraphics[width = 0.6\textwidth]{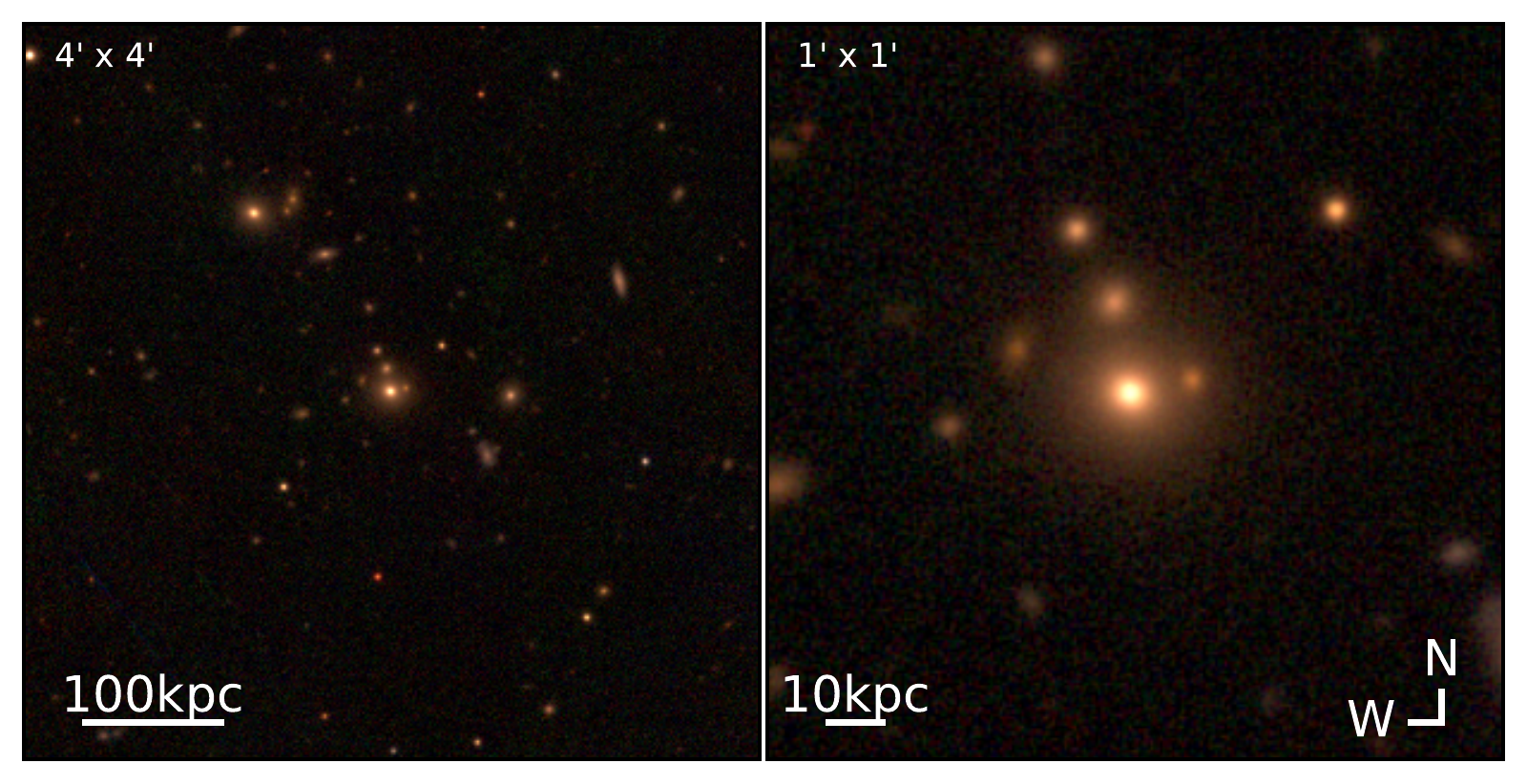}}
{\includegraphics[width = 0.4\textwidth]{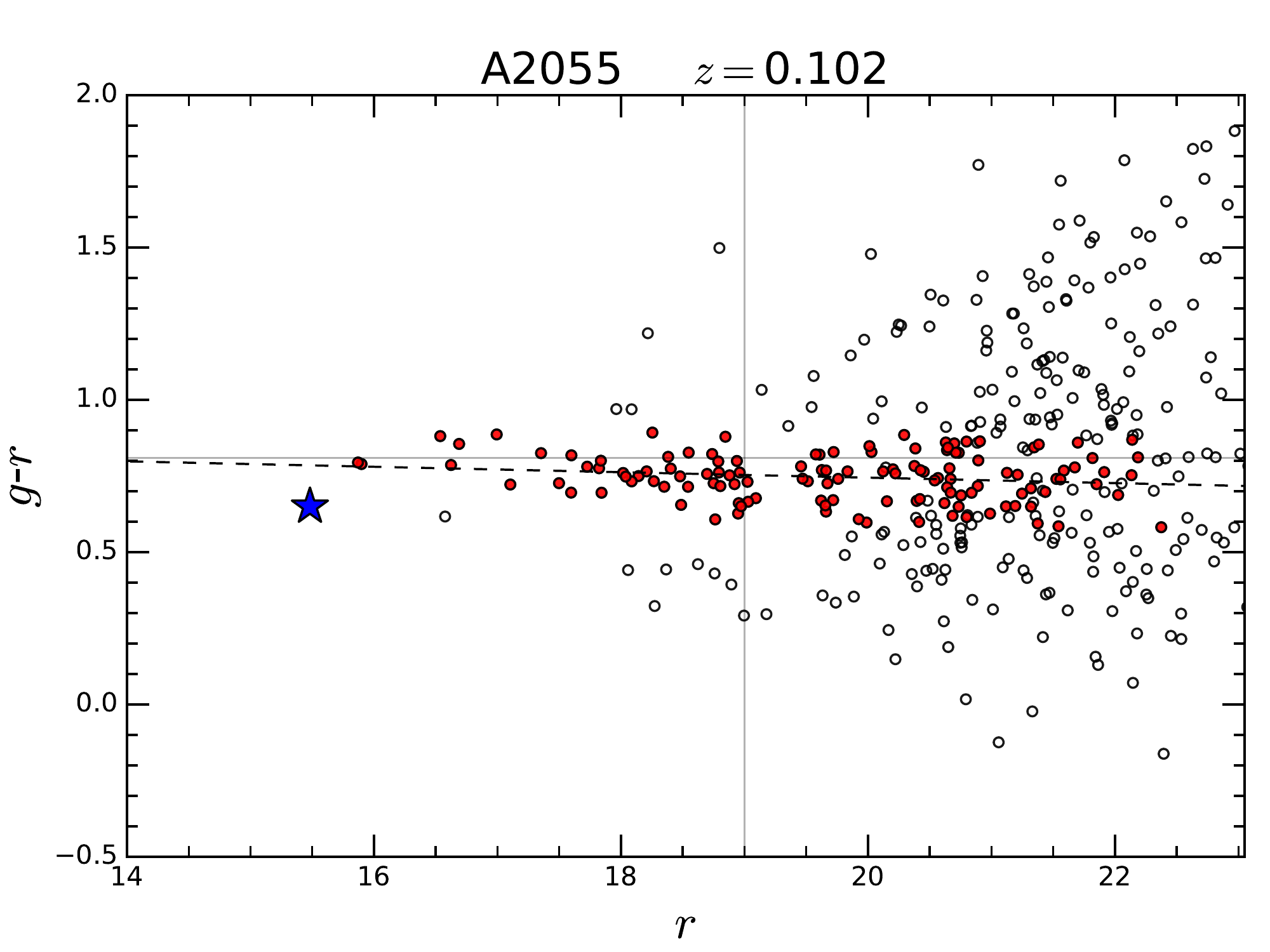}}
{\includegraphics[width = 0.6\textwidth]{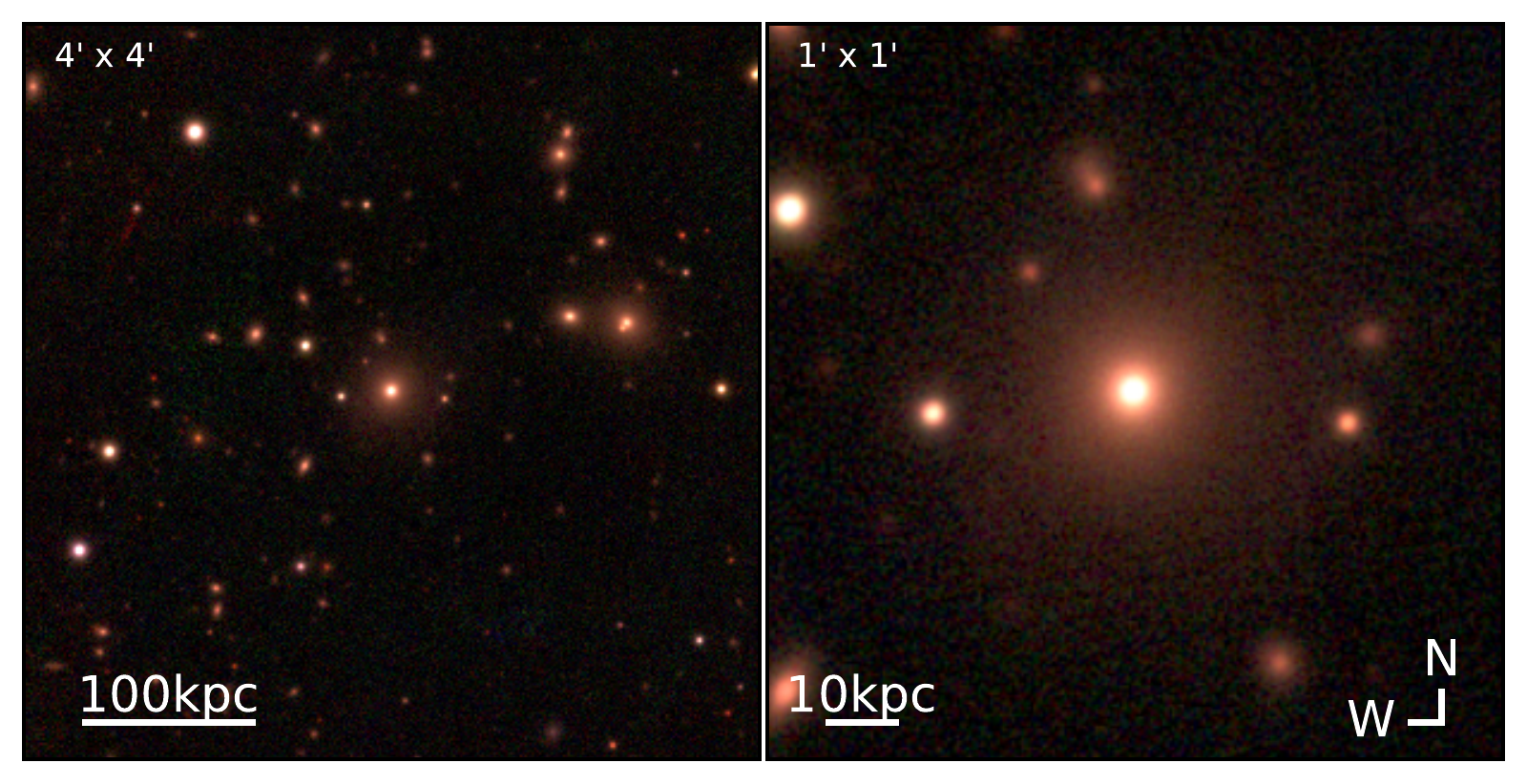}}
{\includegraphics[width = 0.4\textwidth]{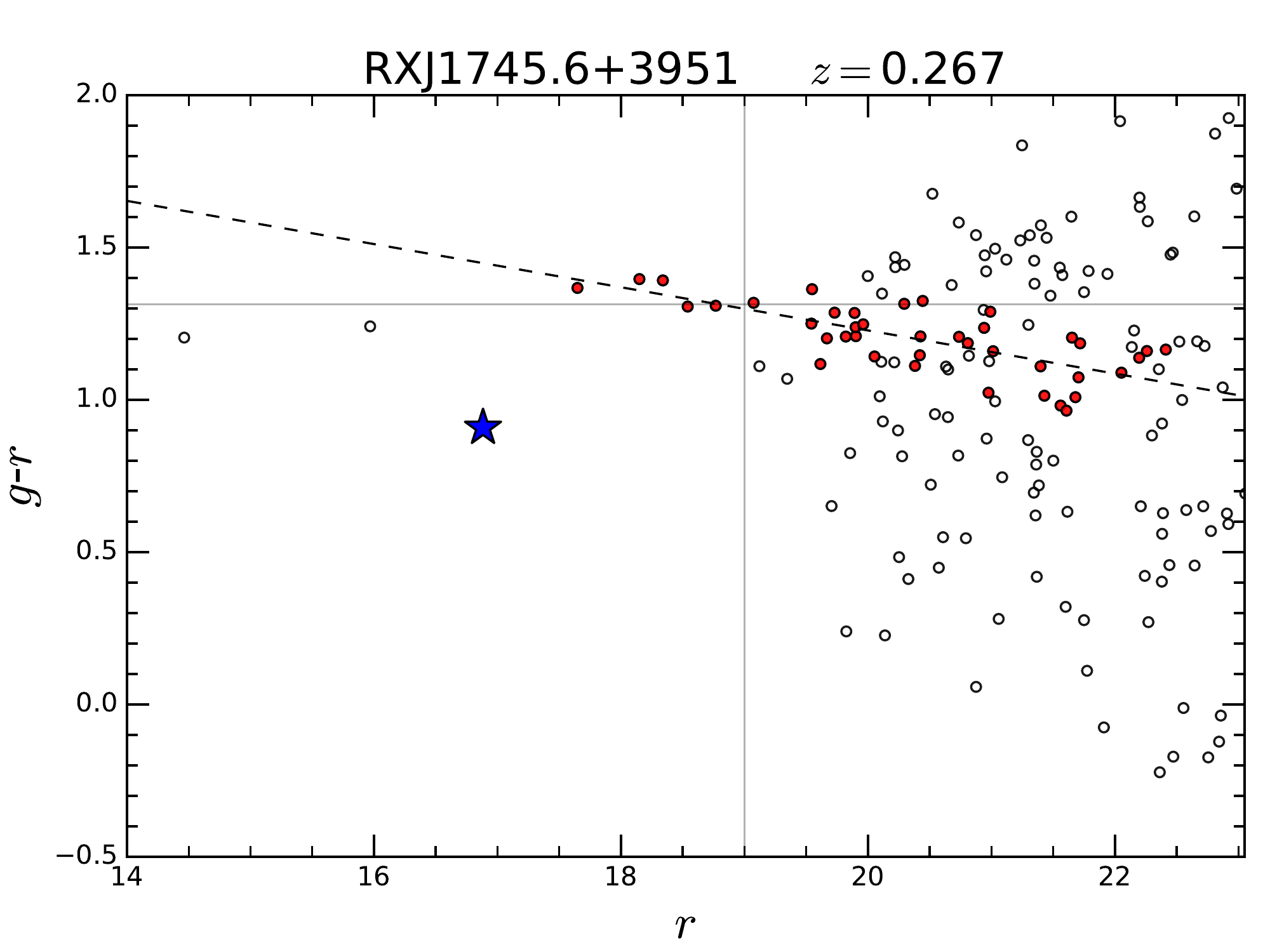}}
{\includegraphics[width = 0.6\textwidth]{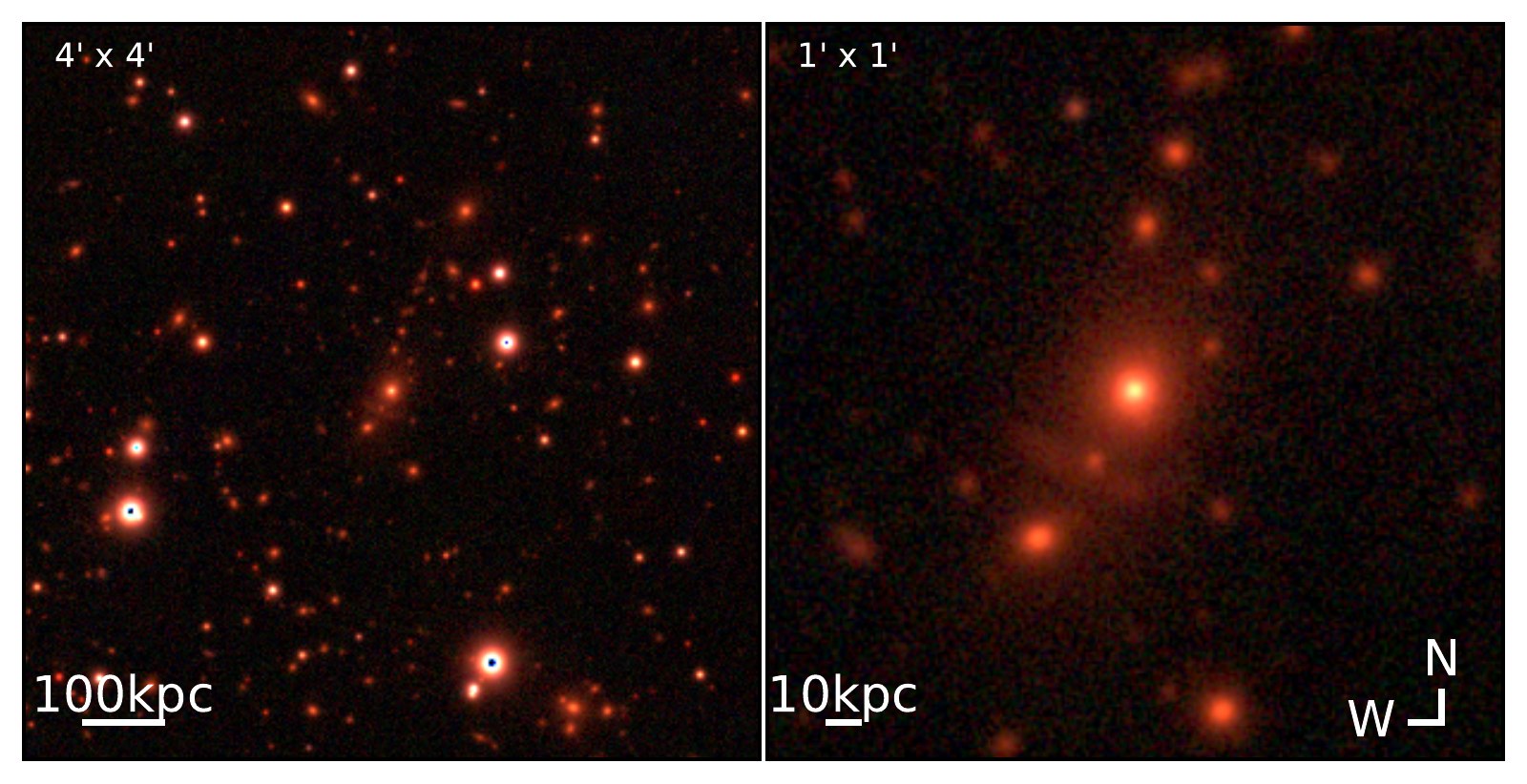}}
\end{minipage}
\contcaption{}
    \label{fig:CMR2}
\end{figure*}

\begin{figure*}
\begin{minipage}{\textwidth}
{\includegraphics[width = 0.4\textwidth]{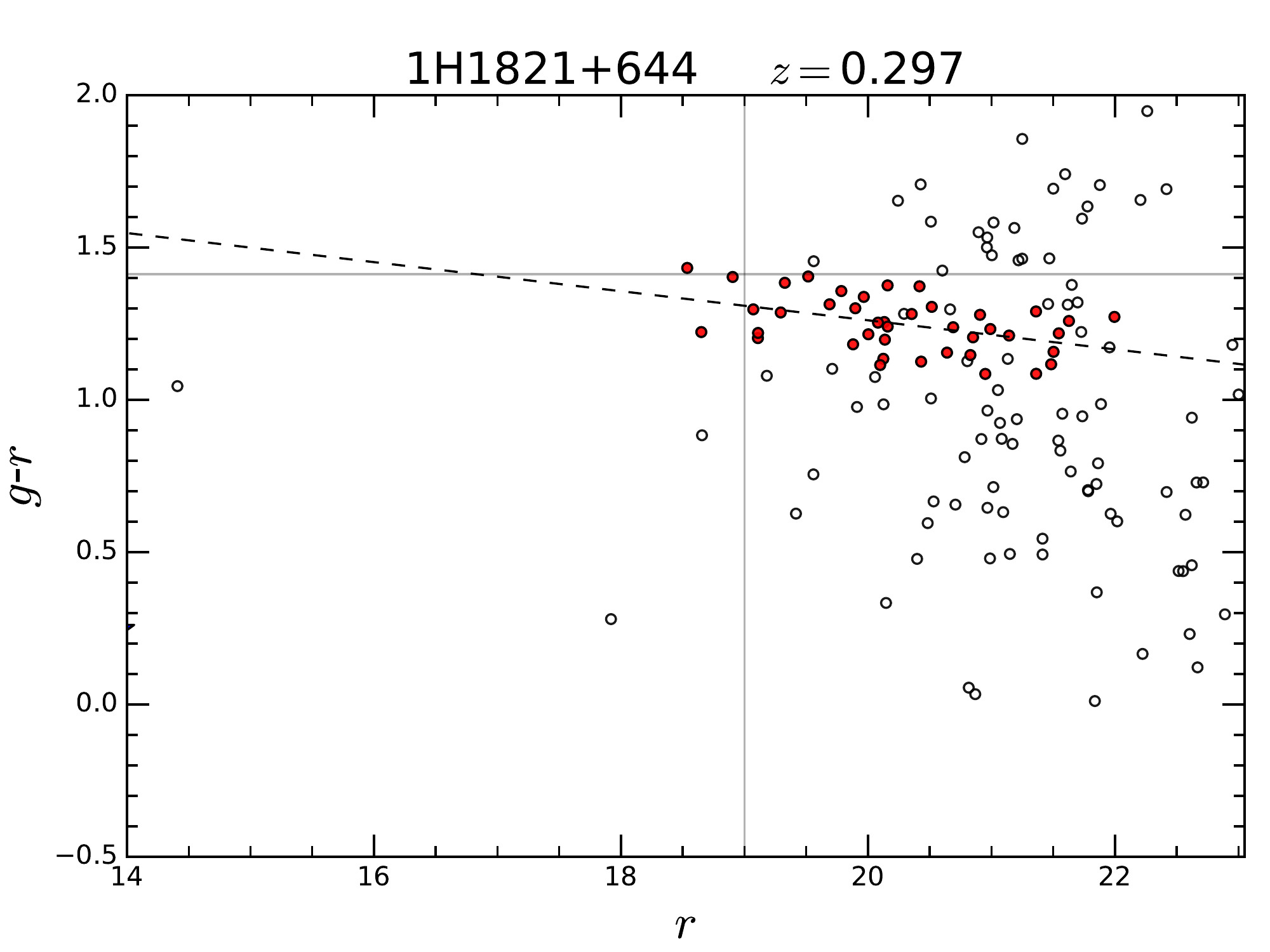}}
{\includegraphics[width = 0.6\textwidth]{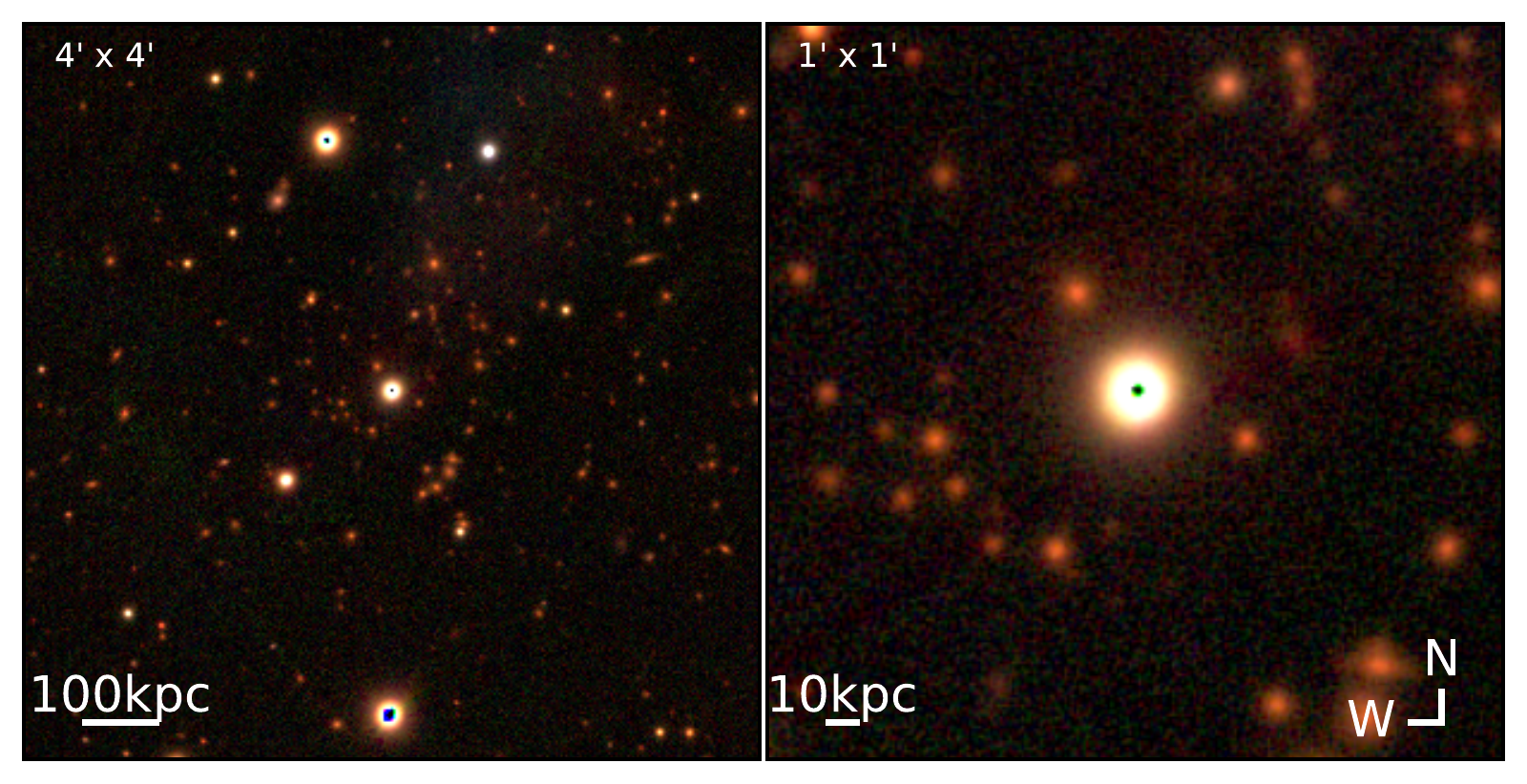}}
{\includegraphics[width = 0.4\textwidth]{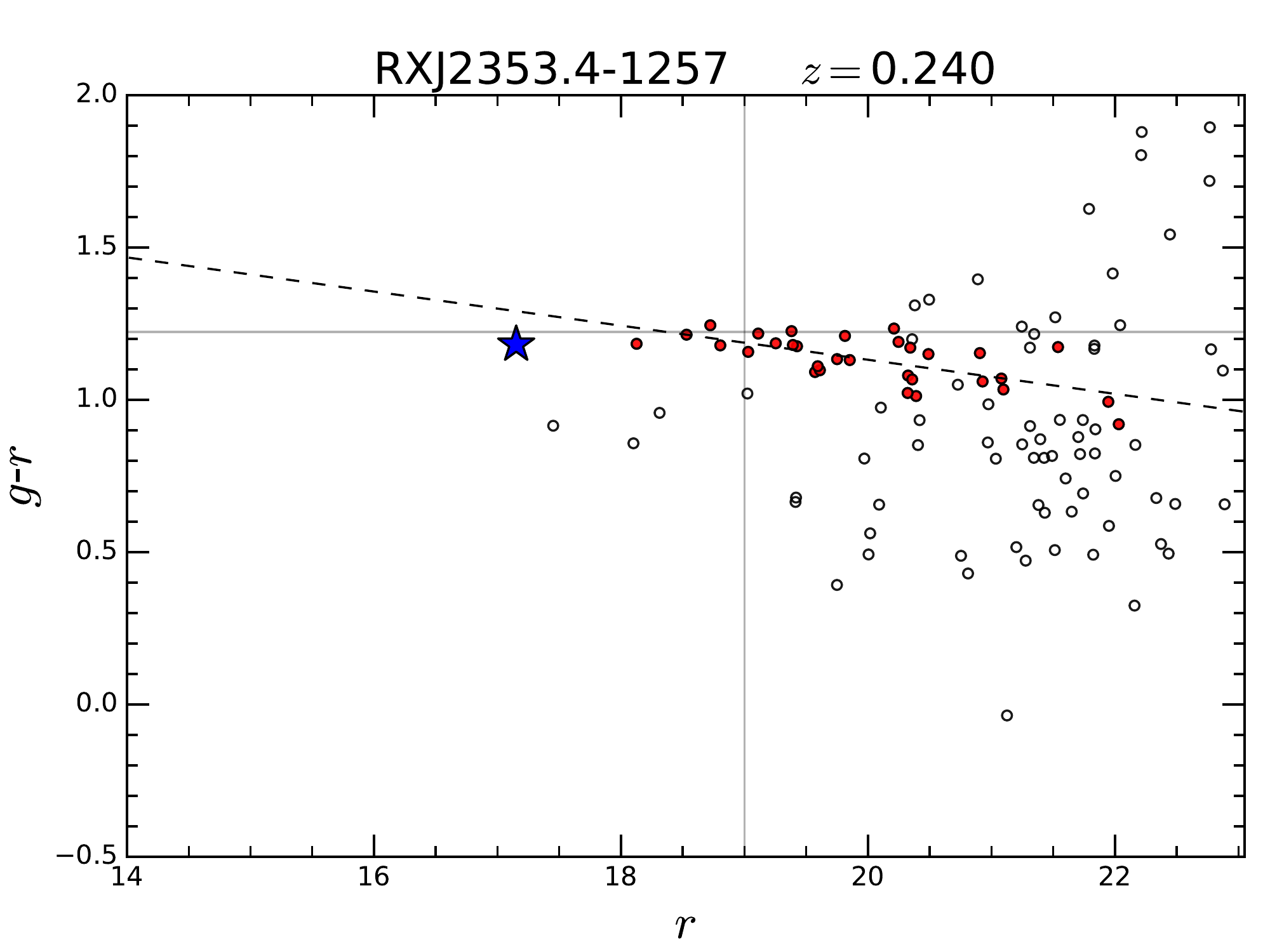}}
{\includegraphics[width = 0.6\textwidth]{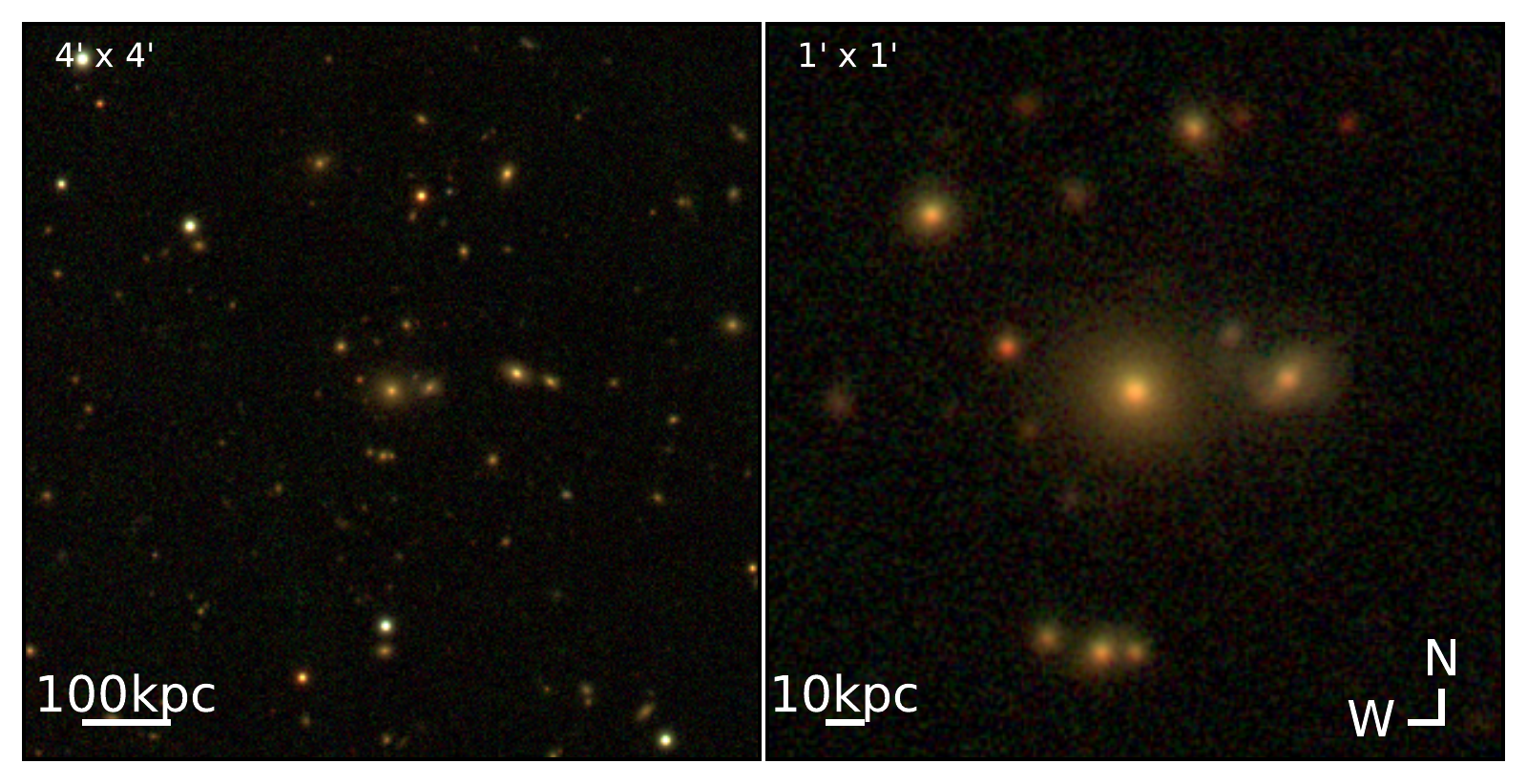}}

\end{minipage}
\contcaption{}
    \label{fig:CMR3}
\end{figure*}


\bsp	
\label{lastpage}
\end{document}